\documentclass[twocolumn,usenames,dvipsnames]{aastex631} 
\usepackage{epsfig,color}
\usepackage{rotating,ulem}
\usepackage{url}

\newcommand{\msun}{\ensuremath{\mathrm{M}_\odot}}

\newcommand{\swift}{\textit{Swift}}
\newcommand{\chandra}{\textit{Chandra}}

\newcommand\HeII{$\textrm{He}\,\scriptstyle\mathrm{II}$}
\newcommand\OII{$\textrm{O}\,\scriptstyle\mathrm{II}$}
\newcommand{\degree}{$^{\circ}$}

\received{2022 October 25}
\revised{\today}
\submitjournal{ApJ}

\shorttitle{LIGHETR} 
\shortauthors{Rosell et al.}

\begin{document}

\title{The LIGO HET Response (LIGHETR) Project to Discover and Spectroscopically Follow Optical Transients Associated with Neutron Star Mergers}

\author[0000-0003-0416-9818]{M. J. Bustamante-Rosell}
\affiliation{Department of Physics, University of Texas at Austin, Austin, TX}
\affiliation{Department of Astronomy, University of California, Santa Cruz, CA}

\author[0000-0003-2307-0629]{Greg Zeimann}
\affiliation{Hobby Eberly Telescope, University of Texas, Austin, Austin, TX, 78712, USA}

\author[0000-0003-1349-6538]{J.\ Craig Wheeler}
\affiliation{Department of Astronomy, University of Texas at Austin, Austin, TX}

\author[0000-0002-8433-8185]{Karl Gebhardt}
\affiliation{Department of Astronomy, University of Texas at Austin, Austin, TX}

\author[0000-0002-7453-6372]{Aaron Zimmerman}
\affiliation{Department of Physics, University of Texas at Austin, Austin, TX}

\author[0000-0003-2624-0056]{Chris Fryer}
\affiliation{Los Alamos National Laboratory, Los Alamos, NM}

\author[0000-0003-4156-5342]{Oleg Korobkin}
\affiliation{Los Alamos National Laboratory, Los Alamos, NM}

\author{Richard Matzner}
\affiliation{Department of Physics, University of Texas at Austin, Austin, TX}

\author[0000-0002-5814-4061]{V. Ashley Villar}
\affiliation{Department of Astronomy and Astrophysics, Penn State University, College Park, PA}

\author[0000-0002-0840-6940]{S. Karthik Yadavalli}
\affiliation{Department of Astronomy and Astrophysics, Penn State University, College Park, PA}

\author[0000-0002-9886-2834]{Kaylee M. de Soto}
\affiliation{Department of Astronomy and Astrophysics, Penn State University, College Park, PA}

\author[0000-0003-0509-2656]{Matthew Shetrone}
\affiliation{University of California Observatories, Santa Cruz, CA 95064}

\author[0000-0001-9165-8905]{Steven Janowiecki}
\affiliation{McDonald Observatory, University of Texas at Austin, Austin, TX}

\author{Pawan Kumar}
\affiliation{Department of Astronomy, University of Texas at Austin, Austin, TX}

\author[0000-0003-4897-7833]{David Pooley}
\affiliation{Department of Physics and Astronomy, Trinity University, San Antonio, TX}
\affiliation{Eureka Scientific, Inc.}

\author[0000-0002-0977-1974]{Benjamin P. Thomas}
\affiliation{Department of Astronomy, University of Texas at Austin, Austin, TX}

\author[0000-0001-5403-3762]{Hsin-Yu Chen}
\affiliation{Department of Physics, University of Texas at Austin, Austin, TX}

\author[0000-0001-7092-9374]{Lifan Wang}
\affiliation{Department of Astronomy, Texas A\&M University, College Station, Texas}

\author[0000-0001-8764-7832]{Jozsef Vink{\'o}}
\affiliation{Department of Astronomy, University of Texas at Austin, Austin, TX}
\affiliation{Konkoly Observatory, Budapest, Hungary}
\affiliation{Department of Optics and Quantum Electronics, University of Szeged, Hungary}

\author[0000-0003-4102-380X]{David J. Sand}
\affiliation{Department of Astronomy, University of Arizona, Tucson, AZ}

\author[0000-0003-3265-4079]{Ryan Wollaeger}
\affiliation{Los Alamos National Laboratory, Los Alamos, NM}

\author[0000-0002-0672-4945]{Frederic V. Hessman}
\affiliation{ Institut f\"ur Astrophysik und Geophysik, University of G{\"o}ttingen,  
Germany}

\author[0000-0001-5538-2614]{Kristen B. McQuinn}
\affiliation{Department of Physics and Astronomy, Rutgers University, Piscataway, NJ}

\correspondingauthor{M. J. B. Rosell}
\email{majoburo@gmail.com}

\begin{abstract}
The LIGO HET Response (LIGHETR) project is an enterprise to follow up optical transients (OT) discovered as gravitational wave merger sources by the LIGO/Virgo collaboration (LVC). Early spectroscopy has the potential to constrain crucial parameters such as the aspect angle. The LIGHETR collaboration also includes the capacity to model the spectroscopic evolution of mergers to facilitate a real-time direct comparison of models with our data. The principal facility is the Hobby-Eberly Telescope. LIGHETR uses the massively-replicated VIRUS array of spectrographs to search for associated OTs and obtain early blue spectra and in a complementary role, the low-resolution LRS-2 spectrograph is used to obtain spectra of viable candidates as well as a densely-sampled series of spectra of true counterparts. Once an OT is identified, the anticipated cadence of spectra would match or considerably exceed anything achieved for GW170817 = AT2017gfo for which there were no spectra in the first 12 hours and thereafter only roughly once daily. We describe special HET-specific software written to facilitate the program and attempts to determine the flux limits to undetected sources. We also describe our campaign to follow up OT candidates during the third observational campaign of the LIGO and Virgo Scientific Collaborations. We obtained VIRUS spectroscopy of candidate galaxy hosts for 5 LVC gravitational wave events and LRS-2 spectra of one candidate for the OT associated with S190901ap. We identified that candidate, ZTF19abvionh = AT2019pip, as a possible Wolf-Rayet star in an otherwise unrecognized nearby dwarf galaxy.  

\end{abstract}

\section{Introduction}
\label{sec:intro}

The first gravitational-wave event discovered by LIGO, the merger of two black holes in a binary system (BBH), opened an exciting new vista in multi-messenger astronomy \citep{2016ApJ...818L..22A} 

Utilizing this new window into the Universe is one of the most exciting prospects in astrophysics. Optical astronomers and observers in other electromagnetic bands were invited to this arena with the advent of GW170817, the result of the inspiral and merger of two neutrons stars (BNS), and its gamma-ray (GW170817A) and optical (AT2017gfo) counterparts. \citep{2017PhRvL.119p1101A,2017ApJ...848L..12A, coulter17,2017Sci...358.1583K, 2017ApJ...848L..24V, 2017ApJ...848L..25H, 2017Natur.551...80K, 2017Sci...358.1559K, 2017ApJ...848L..34M, 2017ApJ...848L..20M, 2017ApJ...848L..18N, 2017ApJ...848L..27T, villar2017combined, troja17, hallinan17, evans17, lyman18, 2018Natur.554..207M, 2018ApJ...856L..18M, 2018ApJ...859L..23P, 2019MNRAS.483.1912P, troja19, lamb19, 2019ApJ...886L..17H, Gillanders22, Kilpatrick22, Hajela22}. AT2017gfo corresponded to a kilonova, a transient about 1000 times brighter than a classical nova \citep{li98,2010MNRAS.406.2650M}.  

Many groups have reported their techniques for searching for optical transients (OT) associated with gravitational wave events in the third LVC run that began on April 1, 2019 \citep[O3;][]{Gomez19, Hosseinzadeh19, Lundquist19, Yang19, Gompertz20, ackley20, kasliwal20, Chang21, Kim21, Paterson21, becerra21, oates21, dejaeger22, Rastinejad22}. Here we report on the LIGO HET Response (LIGHETR) project that employed spectrographs on the Hobby-Eberly Telescope (HET) at McDonald Observatory to conduct a spectroscopic search and follow-up during O3.

The first spectra were taken 0.88 and 1.84 hours after the detection of the OT of AT2017gfo \citep{2017Sci...358.1574S}, fully 12 hours after the original LVC GW signal. The next few spectra were obtained at $\sim$ 1.18 and 1.5 days \citep{2017ApJ...848L..32M, 2017Natur.551...67P, 2017ApJ...848L..18N, 2018MNRAS.474L..71B} and then a few at daily intervals after that, including VLT/X-shooter data extending into the infrared \citep{2017Natur.551...67P}. The target varied rapidly over that timescale and was unobservable spectroscopically after about 10 days. \citet{2017Natur.551...67P} argue for the detection of shallow, broad features at about 8100~\AA\ and 12,300~\AA\ in a spectrum obtained at 1.5 days and several subsequent spectra. \citet{watson} suggest that the former may be evidence of the neutron-capture element strontium. \citet{perego22} and \citet{tarumi23} explore the possibility that the feature is \ion{He}{1} $\lambda 10830$ and stress the possible role of non-LTE effects in differentiating strontium from helium.

Although it was fortuitously nearby, practical factors prevented a higher cadence of spectroscopy of AT2017gfo. The southern location at a declination of -17\degree 51' meant that it was primarily visible in the south and only for about 2 hours per night. A delay in the release of the refined location meant that it was too late to observe from Africa and too early for Chile. Once the OT was accessible in Chile, discovery came quickly and spectroscopy shortly thereafter. The result was an inadequate cadence for such a rapidly-changing event. We were left to wonder about the nature of earlier spectra. 

LIGHETR is a program designed to complement the global effort to obtain spectroscopy of merger optical transient (OT) components. 

The Hobby-Eberly Telescope (HET) with the massively-replicated wide-field VIRUS IFU spectrographs \citep{2018SPIE10700E..0PH,2018SPIE10702E..1KH,hill21,gebhardt21} and the Marcario Low-Resolution Spectrograph (LRS2) \citep[][Hill, et al. 2023, in preparation]{2016SPIE.9908E..4CC,1998SPIE.3355..375H} can play a significant role in multi-messenger astronomy since 1) the queue-scheduled HET is designed to respond quickly to events, 2) the blue sensitivity of VIRUS that extends to 3500~\AA\ can be an important discriminant for various models of the mergers, 3) VIRUS provides the largest sky area coverage of any spectrograph by a factor of 70, 4) the location of GW170817 at the edge of a nearby galaxy highlights the advantage of using an array like VIRUS and 5) LRS2 can provide frequent wide wavelength spectroscopic coverage of these rapidly-evolving events.

With the HET, we can, in principle, discover the OT, get the first spectrum, and then a dense time series of spectra. Ideally, we could obtain several spectra per night, depending on the declination, and maintain this schedule until the OT fades below detectability, in roughly a week. In this time, under ideal circumstances, we could obtain of order 20 spectra and capture in detail the structure and evolution of these incredible events. O3 was a resounding success for the detection of gravitational wave signals from compact star mergers, but disappointing in the lack of observable OTs. Here we describe our attempts to spectroscopically detect six merger events. We failed to see any OT associated with a GW event, but did reveal an interesting object, ZTF19abvionh, that is of intrinsic interest for its own sake.

Section \ref{sec:observations} gives an outline of the LIGHETR observational program. Section \ref{sec:results} presents some results of the program in the first portion of O3 that spanned April 1 to September 31, 2019. Section \ref{sec:S190901ap} presents our follow up of ZTF candidate ZTF19abvionh that proved to be interesting, but not a merger counterpart. Section \ref{sec:assess} explores attempts to quantify our ability to assess the upper limits of undetected sources. Section \ref{sec:models} presents models that illustrate the potential power of early, frequent, spectral observations. Section \ref{sec:future} summarizes the results to date and our future plans.
Appendix \ref{appsec:comp} gives a table of compositions used in kilonovae light curve and spectral simulations, Appendix \ref{appsec:alert} gives some details of LIGHETR operating procedures, and Appendix \ref{appsec:time} presents a notional timetable for responding to an LVC event and initiating a spectroscopic search and followup.

\section{Observational Program}
\label{sec:observations}

LIGHETR was designed to observe about one LVC event per trimester during O3 using the VIRUS IFU array spectroscopically to search for and perhaps discover the OT and to use the LRS2 low-resolution spectrograph to do intensive, high-cadence spectroscopic followup. 
LIGHETR was prepared rapidly to generate the appropriate target information from a trigger and work with the HET staff in order to implement the search and subsequent spectroscopic monitoring as rapidly as possible. 
Our observational strategy involved digesting public alerts with our custom software {\sc diagnosis} and assessing if the candidate is visible to HET. After further human vetting of the candidate, the VIRUS array was utilized to take spectra over a wide field of view centered on the candidate. We have the capacity to carry out further intensive spectroscopic followup with the LRS2 instrument. LIGHETR has the capability to be on target 20 minutes after an alert, achieving dense spectral coverage of the target. LIGHETR can, in principle, observe in bright time since the spectrographs are always mounted. 

The LVC alerts give approximate locations, distances, and estimates of the nature of the merger: BNS, BBH, or NSBH. The LIGHETR program was triggered if an appropriate event, BNS or NSBH, fell within the RA and DEC attainable with HET at that epoch and promised a reasonably bright OT. 

Details of how our program was implemented are given below. A proposed timetable for the LIGHETR response in O3 is given in Appendix \ref{appsec:time}

\vfill\null

\subsection{The Instruments}
\label{instrument}

The Hobby-Ebberly Telescope\footnote{
HET is a collaboration of the University of Texas at Austin, Pennsylvania State University, Stanford University, Georg-AugustUniversität, Göttingen, and Ludwig-Maximillians-Universität, Munich
} (HET) is equipped with a 11 m spherical primary mirror made of hexagonal segments fixed at a zenith angle of 35\degree \citep{hill21}. 
It can be moved in azimuth, in such way that over a day, it can cover $70\%$ of the sky visible at McDonald Observatory.
The pupil, which is 10.0 m in diameter at the center of the track, can be moved to follow an object for different lengths of time depending on the declination.
This tracking varies from 40 minutes at the lowest point (at $\delta = -10.3$\degree) to 2.8 hours at the highest (at $\delta = +71.6$\degree) with a maximum of 5 hours at 65\degree. 

Mounted on the HET are two low-resolution integral field unit (IFU) instruments: VIRUS and LRS2. VIRUS is made up of 156 spectrographs, each fed by a fiber-integral unit, having an overall total fill factor of 1/4.6. Each unit consists of 248 fibers with an individual 1/3 fill-factor, filled out by a three-point dither pattern, moving approximately ~1.5\arcsec\ between each position. While this dither fills in the gaps between fibers in each IFU, it does not fill in the gaps between IFUs. Each fiber has a diameter of 1.5\arcsec\  and each spectrograph covers an area of 50\arcsec\ $\times$ 50\arcsec. The VIRUS wavelength coverage goes deeply in the blue, ranging from $350 < \lambda\: (\mathrm{nm}) < 500$ at a resolving power of $R = \lambda / \delta\lambda \approx 700$. 
LRS2, on the other hand, has a much smaller field-of-view (FoV) of 12\arcsec $\times$ 6\arcsec, with a higher resolution and broader wavelength coverage, spread between two independent fiber-fed dual-channel spectrographs. The blue spectrograph, LRS2-B, with $R = \lambda / \delta\lambda $ of 1900 and 1100 (for the respective channels) covering $370 < \lambda\: (\mathrm{nm}) < 700$ and the red spectrograph, LRS2-R, with $R = \lambda / \delta\lambda $ of 1800 covering $650 < \lambda\: (\mathrm{nm}) < 1050$.
The response functions of LRS2 will be presented in Hill et al. (2023, in prep).

Typical observing conditions at HET have a median seeing of 1.7\arcsec. Under such conditions, a baseline 20 minute observation with three dithered exposures of 360 s has a line sensitivity at 5 $\sigma$ of $\sim 6\times10^{-17}$ erg cm$^{-2}$ s$^{-1}$. The fill factor of VIRUS of 1/4.6 limits its capacity to get several galaxies in one dither pattern. This is why we do targeted galaxy follow-up, placing individual galaxies at the center of an individual IFU.

\subsection{Alert}
\label{alert}

In O3, if automated vetting detected a potential merger event, LVC then sent an automated public GCN Preliminary Alert within about 10 seconds to 1 minute of the detection. Each alert was subject to human vetting and either an Initial Alert or a retraction was sent on a timescale of tens of minutes to an hour. The GCN Preliminary Alert pointed to the LVC sky map that gave the 3D probability distribution of the source. 

Both due to the rapidly varying nature of kilonovae and the fact that HET is a fixed zenith telescope, which gives it limited tracking capability, quick knowledge of when the region of highest probability for a given event would be visible required a custom alert system. 
With this in mind, we created {\sc diagnosis}\footnote{https://github.com/Majoburo/Diagnosis}, a low-latency alert system and observation planner for HET. {\sc diagnosis} takes into account the geometry of the HET pupil and its position at any given point in time to submit an observation request to the HET Resident Astronomers and to alert the LIGHETR team. 

\begin{figure*}[!ht]
   \begin{center}
\includegraphics[width=\textwidth]{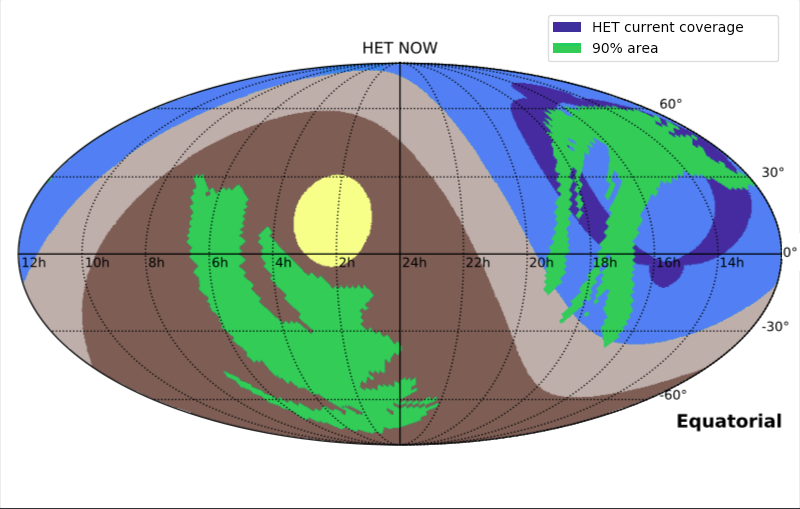}
      \caption{
The sky map produced by using LVC localization data for the merger event S190425z and the HET track corresponding to the epoch of the alert. The LVC location of this alert is given by the 
green bands. The region accessible to the HET above DEC = -10 degrees and below +71 degrees is given by the dark blue 
band on the right. The lighter blue 
represents air mass less than 2.5. The location within 18 degrees of the Sun is given by the large yellow 
patch. The lighter brown 
area is air mass greater than 2.5 and the dark brown 
region on the left represents the portion of the sky below the horizon. The HET track overlapped with the large LVC sky map at the time of this alert. This figure provides a visual aid for observers and illustrates the technique by which we decided whether to trigger on a given LVC alert and to construct a prioritized list of galaxies to be searched for a new optical transient. 
         }\label{ligohetmap}
 \end{center}
\end{figure*}

{\sc diagnosis} continuously listened to the Gamma-ray Coordinates Network/
Transient Astronomy Network (GCN/TAN)\footnote{https://gcn.gsfc.nasa.gov/} for alerts on gravitational wave events. When triggered, if the event was likely to have an electromagnetic counterpart\footnote{If it was likely to be a BNS, NSBH, or in the mass gap (one or more of the binary objects is in the range of 3 to 5 \msun). We did not pursue any BBH candidates in O3.}, {\sc diagnosis} downloaded the associated skymap, identified if and when the 90\% confidence region fell within the HET pupil, and if so, informed the observers. As an example, Figure~\ref{ligohetmap} shows the sky map produced by {\sc diagnosis} for the merger event S190425z, which turned out to be the second confident BNS detection by the LVC \citep{Antier20,2019ApJ...885L..19C}. 
{\sc diagnosis} also queried a galaxy catalog for galaxies within the observable 90\% probability region, organized them by probability, gave their local sidereal times (LSTs) to start observations and made a submission file for HET observations. An example of a graphic representation of the notice sent to the observers is in Figure~\ref{alert} in Appendix \ref{appsec:alert}.

\subsection{Galaxy Catalog} 
\label{catalog}

\begin{figure*}
	\centering
	\includegraphics[width=\textwidth]{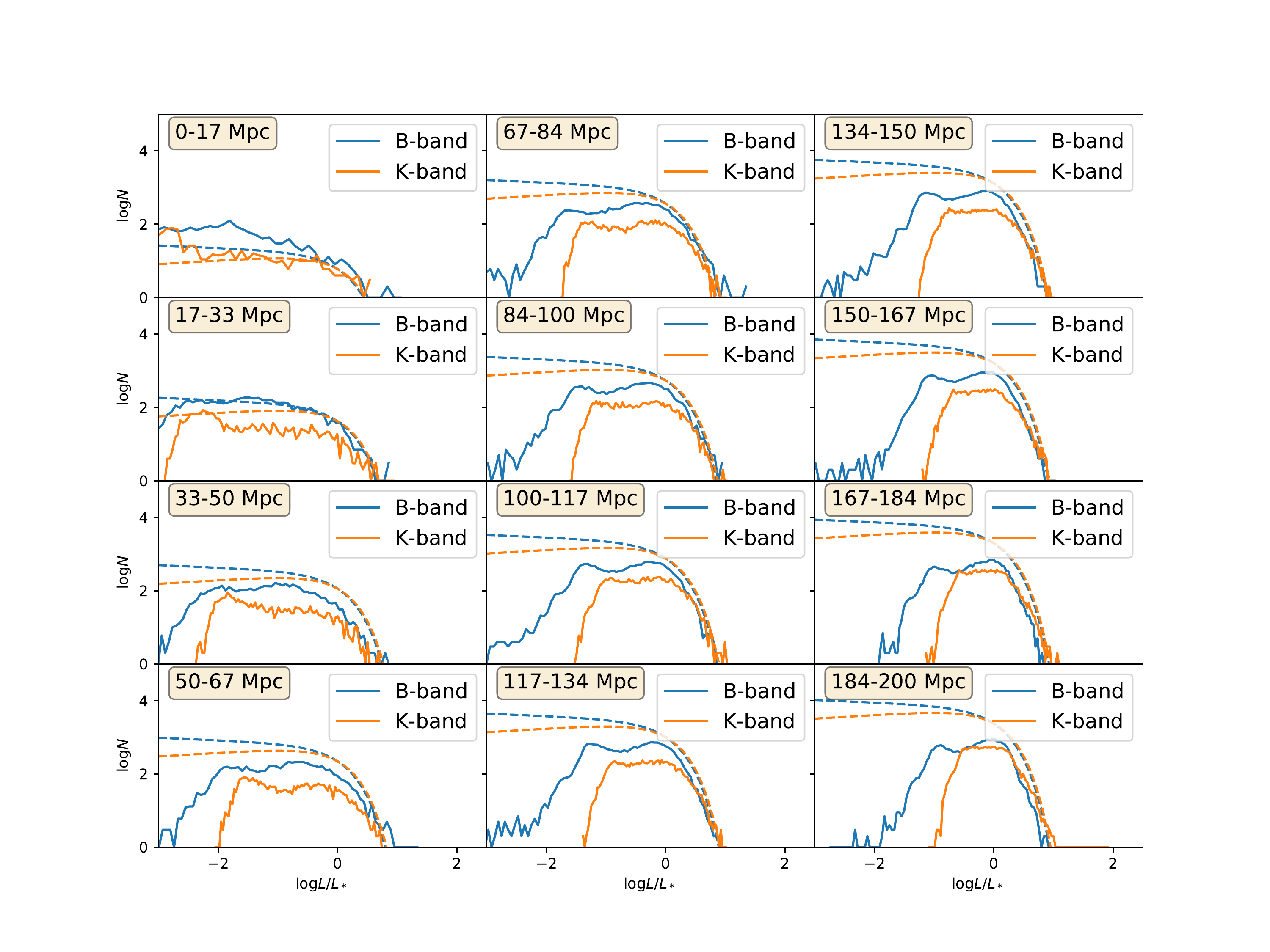}
	\caption{Completeness of the GLADE catalog in the B and K bands for different redshift ranges. The solid lines are luminosity histograms of GLADE galaxies within different distance shells in terms of their measured B-band and K-band luminosities. Dotted lines are our expectations for complete catalogues based on Schechter function measurements for each band. The GLADE catalog is designed to be more complete in the B-band, but we use the K-band since we aim to use the luminosity as an indicator of mass.  }
	\label{fig:completeness}
\end{figure*}

We used version 2.3 of the GLADE galaxy catalog \citep{Dalya2018}. The GLADE catalogue contains around 3.26 million objects (149 globular clusters, 297014 quasars, and 2965717 galaxies). It was constructed combining data from 5 astronomical catalogues: GWGC \citep{White2011}, 2MPZ \citep{Bilicki2013}, 2MASS  XSC \citep{Skrutskie}, 
HyperLEDA \citep{Makarov2014}, and SDSS-DR12Q \citep{Paris2017}. The catalog is complete up to $d_L = 37^{+3}_{-4}$ Mpc in terms of cumulative B-band luminosity out to such distance, decreasing to 50\% completeness at $d_L$ of 91 Mpc. 
Figure~\ref{fig:completeness} shows the catalog's completeness in both B and K bands as a function of luminosity for different redshift bins. The expected detection depth for LVC merger events in O3 was 100 - 140 Mpc. This substantially exceeds the completeness depth of the Glade catalog, but the catalog contains some galaxies extending to the LVC limit.

\begin{figure*}[ht]
	\centering
	\includegraphics[width=\textwidth]{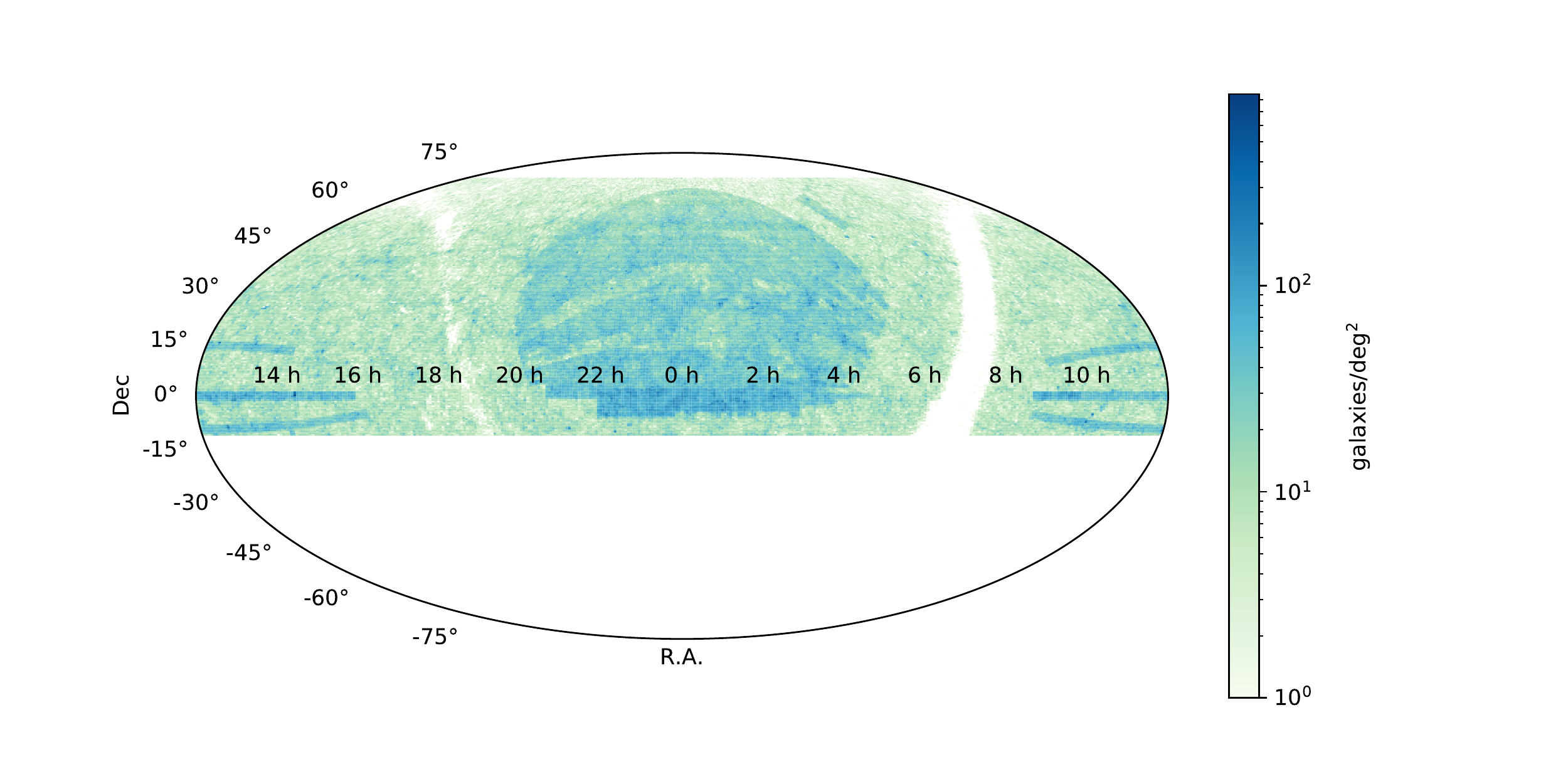}
	\caption{Galaxies contained within the GLADE catalog that are visible from HET.}   \label{fig:gladeHET}
\end{figure*}

We constrained the selection of galaxies in the catalog to those visible by HET and those for which we could estimate their mass, that is, galaxies within declinations of 71.6$^o$ to -10.3$^o$ and with luminosity distances and K-band magnitude measurements. This cut our list of objects to 19\% of the original list, leaving only galaxies. Figure~\ref{fig:gladeHET} shows the number density in the sky of our selection of GLADE galaxies.

\subsection{Galaxy Sorting} 
\label{sort}

\citet{2016ApJ...820..136G} estimated that for bright galaxies contributing $\sim 50\%$ of the light in a given area of the sky, there will be $\sim 20$ galaxies inside a typical LVC error box that are also consistent with LVC distance estimates. In order to rank the most probable galaxies, we weighed the localization probability given by the LVC \citep{Singer2016} by the galaxy mass, which has been shown to be linearly proportional to the K-band luminosity \citep{Kauffmann_1998}. Previous work took B-band luminosity as a weight factor on the galaxy selection process \citep{Arcavi_2017,Yang19}.
Even though the GLADE catalog we used is more complete in terms of galaxies with B-band luminosities, these luminosities have the problem of being affected by star formation history and dust extinction, which makes them an unreliable indicator for mass.
Using the B-band without such considerations would effectively establish a preference towards star-forming galaxies. Both simulations and measurements have shown that short soft gamma-ray bursts (SGRBs) are produced by BNS mergers \citep{Wiggins_2018}. GW170817 confirmed this expectation. A recent kilonova candidate was associated with a smaller galaxy \citep{troja22,yang22} and SGRBs are found in a variety of galaxy types \citep{oconnor22}. Using a galaxy ranking based on K-band luminosity might aid in the identification of events like GW170817 but might discriminate other candidates in other host galaxy types. Despite significantly reducing the completeness of our galaxy sample by constraining it to those with K-band luminosities as shown in Figure~\ref{fig:completeness}, we chose galaxy mass and hence K-band luminosity as our weight factor in the localization probability.

The LVC provided a localization probability, $\rho_i$, as a 2D HEALpix skymap.
For each pixel, i, within the skymap, the LVC also included Gaussian parameters
($\sigma_i$; $\mu_i$) for the distance probability. Thus, we took the galaxies within our catalog falling in the 90\% confidence region in terms of the 2D skymap, calculated their distance probability in terms of their catalogued luminosity distances, $r_g$, and multiplied this by their K-band luminosity, $L^g_K$, and 2D localization probability, $\rho_i$, to get their final ranking, 
\begin{equation}
    \label{eq:galaxy_ranking}
    R(r_g,L_K^g) \propto L_K^g \rho_i \frac{1}{\sigma_i} \exp\left[-\frac{\left(r_g - \mu_i\right)^2}{2{\sigma_i}^2}\right] r_g^2 \,.
\end{equation}
This ranking was normalized to add to unity when summing over the most probable 100 galaxies.
\vfill\null

\subsection{Search for the Optical Transient} 
\label{search}

As described above, the probability distributions of the LVC sky maps and HET track maps illustrated in Figure~\ref{ligohetmap} were automatically employed by {\sc diagnosis} to query the {\sc glade} catalog to produce a prioritized list of galaxies to search for the OT. Such galaxy lists were the basis for the VIRUS search. {\sc diagnosis} also generated the Phase II Target Submission List (TSL) for the prioritized list of galaxies. The result was a prioritized list of galaxies in the HET overlap region and within the distance window provided in the LVC alert. We then systematically observed that sample of galaxies with the VIRUS array. Examples of the LIGO alert as processed by {\sc diagnosis} for local redistribution, the galaxy priority list, and the selection of galaxy priorities are given in Appendix \ref{appsec:alert}. Within O3, our observations all fell within the $\sim 10$ most likely galaxies according to our prioritized list.

\begin{figure}[!ht]
   \begin{center}

\includegraphics[width=\linewidth]{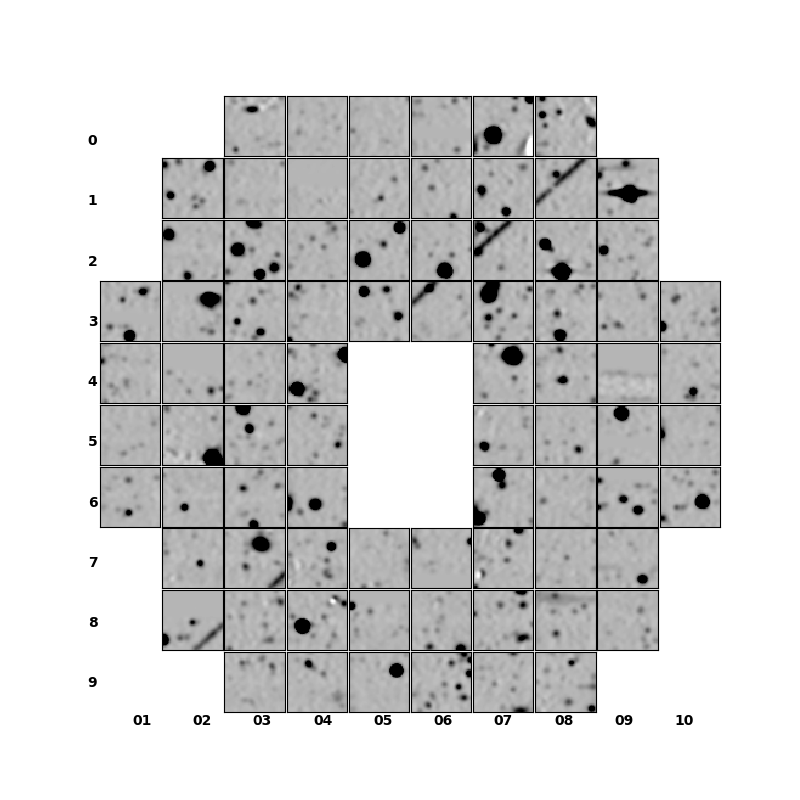}
      \caption{
A sample VIRUS \citep{gebhardt21, hill21} image that would be used in our proposed search for the optical transient associated with a gravitational-wave event. Each IFU unit covers an area of 50\arcsec\ on a side. The VIRUS array automatically produces a zeroth-moment image using the wavelength region 3500-5500 \AA\ and an adopted aperture to make a collapsed image. The whole array is 21\arcmin\ on a side. This image butts the IFUs together preserving their relative position. Each of the 20,000 fibers is positioned to an accuracy of 0.1\arcsec. The faintest objects in the image are g$\sim$22-23. The streak in the upper right is a satellite or asteroid artifact. 
         }\label{fig:hetdex}

 \end{center}
\end{figure}

Once we start the search, a key capacity is to rapidly reduce the data to decide whether there is an OT in a given galaxy or to continue to the next galaxy on the priority list. VIRUS observations always produce a zeroth-moment image as one of the first steps. The VIRUS IFU units do not physically touch one another. The wavelength region of 3500-5500 \AA\ in a selected aperture is used to make a collapsed image, as illustrated in Figure~\ref{fig:hetdex}. 
This initial collapsed image was compared to archival imaging data (specifically from Pan-STARRS), a step which could (but did not in O3) promptly reveal the OT.

During O3, VIRUS spectrograph units were steadily being added to the array. At the time, the array observed about 80 \% of its field that is 21\arcmin\ on a side (Figure~\ref{fig:hetdex}). Each IFU unit covers an area of 50\arcsec\ on a side. Normal VIRUS operation is to do a three-point dither to fill the fiber spaces. We employed this operation since it enhanced our ability to calibrate the spectrum.

For rapid data reduction, we used {\sc remedy} v0.1, a stripped-down version of the LRS2 reduction code, {\sc panacea} \citep{zeimann}\footnote{Panacea v0.1 documentation can be found at: https://github.com/grzeimann/Panacea/blob/master/READMEv0.1.md}. The initial success of {\sc remedy} encouraged an evolved version of the code to become the HET's default pipeline for VIRUS observations. A full description of {\sc remedy} can be found in Zeimann et al. (2023, in prep).  {\sc remedy} has the capability of reducing the full frame or just the data from a dithered IFU exposure of the host galaxy and its vicinity. We then produced a collapsed image from the IFU and a reduced spectrum, all in 30 seconds. 

We pointed a single VIRUS IFU at a target host galaxy while getting spectra of all the other galaxies in the VIRUS field simultaneously. Depending on the target density, circumstances might have allowed us to multiplex, acquiring multiple target galaxies in a single VIRUS field. We note that the merger event GW170817 was about 10\arcsec\ away from its host galaxy. With the 50\arcsec\ square VIRUS IFU field of view, had we pointed at the host, we would easily have picked up the OT and acquired its spectrum (Figure~\ref{fig:hetdex}). 

Each VIRUS set of three dithered exposures of a target galaxy required a total of $\sim 15$ minutes, including shuffling between galaxies. The total time to observe 20 galaxies was thus $\sim$ 5 hours. We were not able to reach this limit in O3 because our targets only became accessible to HET a few hours before sunrise.

\begin{table*}
\caption{Estimated Number of Single Track Spectra Per Night as a Function of Declination and OT Magnitude.}
\centering
\label{tab:spectra}
\begin{tabular}{lccccc}
\hline
\hline
Declination & Track Length & Magnitude & Exposure Time & Number of Single Spectra & Number of B + R Spectra  \\
degrees & minutes & mag & minutes &   &  \\

\hline
\hline

20 & 72 & 17.5 & 20 & 3 & 1  \\
45 & 96 & 17.5 & 20 & 4 & 2  \\                                        70 & 156 & 17.5 & 20 & 7 & 3   \\       

\hline

20 & 72 & 21.0 & 60 & 1 & 0  \\
45 & 96 & 21.0 & 60 & 1 & 0  \\                                       70 & 156 & 21.0 & 60 & 2 & 1   \\  

\hline  
\hline

\end{tabular}
\end{table*}

\subsection{Dense Spectral Followup}
\label{spectra}

Once an OT candidate was identified by our search or others, we planned to implement follow-up spectroscopy with LRS2. 
The spectra of a kilonova are expected at first to change on timescales of hours and later on time scales of days. 
We could, in principle, take spectra for about a week's span with as rapid a cadence as the HET would allow using all the viewable time. 

Table \ref{tab:spectra} presents a short summary of track lengths and number of spectra expected as a function of the brightness and declination of the target. We assume that the total setup time is 10 minutes for a single track following a single object. The actual setup time is often less than 10 minutes, and there is very little setup time, perhaps one or two minutes, to switch from LRS2-B and LRS2-R if both observe the same object. For an OT as bright as AT 2017gfo, $\sim 17.5$ mag in the i band at discovery, we assume 20 minute exposures for each early observation and hence that a single spectrum would require 10+20 = 30 minutes and that a full LRS2-B and LRS2-R spectrum requires 10+20+20 = 50 minutes. For long tracks allowing multiple exposures of the same object, the total exposure would be 10 minutes plus the number of 20 minute intervals that sum to less than the track length. If we elect to get only LRS2-B in the early phases, then we would get roughly twice as many spectra per track than obtaining LRS2-B and LRS2-R spectra, with the caveat that we could only get an even number of B + R spectra utilizing a single track. 

Thus, in an ideal northern declination case $\sim +70^o$ with a track length of $\sim 156$ min and target of brightness $\sim 17.5$ mag, we can expect no more than overhead plus seven consecutive 20 minute shots per track. We could get three full LRS2-B + LRS2-R spectra per track. If the object is at lower declination, $\sim 20^o$, then we can still get three single spectra or two full spectra per track.

Events discovered in O3 were on the average more distant and fainter than AT2017gfo, requiring longer exposures. AT2017gfo was at only $\sim 40$ Mpc. LVK will reach to 160 to 190 Mpc in O4, implying a factor of 16 to 23 in luminosity and 3 to 3.4 mag in brightness. A target in this distance range would be about r = 20.5 to 21, requiring an exposure of about an hour on the HET. A single spectrum including overhead would require 70 minutes. To get both LRS2-B and LRS2-R would require 130 minutes. At this brightness at a declination of $\sim +70^o$ with a track length of $\sim 156$ min and target of brightness $\sim 21$ mag, we might acquire two single spectra per track and one complete LRS2-B + LRS2-R spectrum per track. At declinations $\sim 20^o$ and $\sim 45^o$, we could expect only a single spectrum per track.  

If we assume the target is visible throughout the whole night, once on an East track and once on a West track, a somewhat unlikely circumstance in practice, then we could get about twice as many spectra per night. For a declination of $\sim 70^o$ at $\sim 17.5$ mag, we could get 14 single spectra per night or six complete LRS2-B + LRS2-R spectra. At lower declinations of $\sim 20^o$ and $\sim 45^o$ and $\sim 21$ mag, we would need this special circumstance to acquire LRS2-B on one track and LRS2-R on the other to get even a single complete LRS2-B + LRS2-R spectrum per night.

While somewhat optimistic, this cadence of spectra would match or considerably exceed anything achieved for AT2017gfo. At later epochs, the OT will be dimmer, also calling for longer exposures, but the cadence can be more relaxed. A Timetable for Alert, Search, Dense Spectral Sampling, and Analysis is given in Appendix \ref{appsec:alert}

\section{Results}
\label{sec:results}

Details on the number of exposures on all the events we followed in O3 are summarized in Table \ref{tab:summary}.

\begin{deluxetable*}{c|c|cccc}
	\tablecaption{Summary of LIGHETR Observations}
	\tablewidth{0pt}
	\tablehead{
        \colhead{Date (GMT)} & \colhead{Event} & \colhead{Exposures} & \colhead{Instruments} & \colhead{Searched} & \colhead{Follow-up}
	} 
	\startdata	
    19/04/12 & S190412m & 21 & VIRUS & 7 & 0 \\
    19/04/25 & S190425z & 18 & VIRUS & 6 & 0 \\
    19/04/27 & S190426c & 15 & VIRUS & 5 & 0 \\
    19/08/22 & S190822c\tablenotemark{a} & 9 & VIRUS & 3 & 0 \\ 
    19/08/29 & S190829u\tablenotemark{a} & 6 & VIRUS & 2 & 0 \\
    19/09/03 & S190901ap & 3 & VIRUS & 1 & 0 \\
    19/09/03 & S190901ap & 3 & VIRUS & 0 & 1 \\
    19/09/06 & S190901ap & 5 & VIRUS, LRS2 & 0 & 1 \\
	\enddata
     \tablecomments{Summary of all events we followed. Retracted events are shown in red. 1) Date searched; 2) GCN/TAN event name; 3) Number of Exposures; 4) Number of searched Galaxies; 5) Number of external EM triggers followed.}
     \tablenotetext{a}{Retracted event}
	\label{tab:summary}
\end{deluxetable*}

With all the elements in place for a successful program, we recognized that this is a complex process that had never been employed on the HET. For this reason, we did a dry run during the day with a practice alert trigger. 

No event in O3 revealed an OT to any facility, so we did not employ LRS2 (but see Section \ref{sec:S190510g}).

\subsection{Dress Rehearsal: GW190412m}
\label{sec:S5190412m}

To complement the dry run,  we also planned a full dress rehearsal initial run. For this, we proposed to trigger on the first LVC alert of O3 that was accessible to the HET. The probability was that this first alert would be a BBH merger event and hence most likely not accompanied by an electromagnetic signal. There was a small possibility that the first alert of O3 might correspond to a BNS or NSBH merger and an associated OT. Theoretical modeling and observations hinted that a BBH merger within the gas-dense environment of a supermassive black hole accretion disk might generate an OT \citep{1999ApJ...521..502C, 2017ApJ...835..165B, 2020PhRvL.124y1102G, 2022MNRAS.512.2654P}. In any case, responding to this alert would give us critical feedback that all our interlocking components, hardware, software and human, would function as needed for a merger with an OT.

For our dress rehearsal, we responded to BBH candidate GW5190412m, only the second event of O3 and the first in range of the HET. The LVC alert for GW190412m was ingested by our {\sc diagnosis} pipeline, which created a prioritized list of target galaxies, a corresponding TSL, and alerted our team. We were awakened at 4 am local Texas time and were on the sky and taking data within about 20 minutes. We observed 7 of the top priority list of 13 target galaxies in the overlapped LVC/HET field. We drafted and sent a GCN notice about our observations in real time and submitted it as the Sun rose on the Observatory \citep{roselle19a}. In hindsight, the LVC alert had been issued about two hours earlier than our local alert. This did not impede our observations in practice, and the problem leading to the local delay was addressed and eliminated. We also found that a vestigial instruction in the TSL requested shorter exposures than ideal. This allowed us to make more observations, but with less S/N ratio than preferable. We corrected this in subsequent observations. Despite the expected absence of an OT for S5190412m, the dress rehearsal served its purpose.

\subsection{GW190425z}
\label{sec:S190425z}

The next event with the potential to produce an optical signal was GW190425z with a probability greater than $99\%$ of being a BNS merger. This was the sort of event for which our program was designed. The local alert was released at 4:01 am local time. Our target list of high priority galaxies was loaded into the HET queue at 4:07 am. Unfortunately, the LVC sky map was very large since only LVC Livingston and Virgo were operating. The distance was estimated by LVC to be 155 $\pm$ 44 Mpc, much more distant than GW170817 at 40 $\pm$ 8 Mpc. 
 
Pan-STARRS images of our target galaxies were shared at 4:48 am. We sampled a prioritized list of 4 galaxies from the GLADE catalog that overlapped with the LIGO probability map and the observable pupil of the HET. 
The effective limiting magnitude in the B band was $\sim 22$ magnitudes. 
A GCN was submitted at 6:51 am \citep{roselle19b}. We did not detect an OT, and neither did any other team. Some transient candidates were reported, but they all proved to be more prosaic, mostly supernovae. We elected not to continue our search of the LVC sky map on a second night. LANL team members (CF, OK, RW) worked through this period making models corresponding to the LVC distance (see \S \ref{sec:models}) and even of some of the transient candidates. LIGHETR group member JV and collaborators observed the five galaxies targeted by HET at Konkoly Obervatory. CCD frames were taken with the 0.6/0.9m Schmidt telescope (FoV 70x70 arcmin$^2$, unfiltered, limiting mag ~21.5) and the 0.8m RC telescope (FoV 18x18 arcmin$^2$, g- and r-band, limiting r magnitude $\sim 20.4$). 

\subsection{GW190426c}
\label{sec:S190426c}

The LVC announced an alert on GW190426c at 10:47 am local time the next morning. This event had a 49\% chance of being a BNS but a 13\% chance of being a NSBH merger and a 14\% chance of being noise. The estimated distance was very large, 375 $\pm$ 108 Mpc. The LANL group produced new light curve models based on the estimated distance. Half the HET track was lost to solar occlusion. Target galaxies in the useable track became observable about 4:30 am. After an extensive discussion of the distance, the expected S/N ratio, the Moon location, the expected rarity of targets, the competition from other groups, and the likelihood of tentative (but ultimately irrelevant) OT candidates, we elected to do our VIRUS search on this event. We sampled a prioritized list of 5 galaxies from the GLADE catalog and submitted a GCN at 6:31 am \citep{roselle19c}.

\subsection{S190510g}
\label{sec:S190510g}

The next potentially interesting LIGO event, S190510g, had a probability of 42\% of being a BNS but 48\% of being noise. The distance was also estimated to be rather large, 280 Mpc. After some debate, we passed on this event. 

\subsection{GW190901ap}
\label{sec:S190901ap}

There were no other alerts that required a response from us until GW190901ap, which had an 86\% chance of being a BNS merger. We observed one target galaxy with VIRUS and sent a GCN \citep{roselle19d}. The next morning, the Zwicky Transient Factory (ZTF) announced four possible OT candidates consistent with the redshift estimated by LIGO \citep{kool19}. Three were ruled out by other groups. 

\begin{figure*}[!ht]
   \begin{center}
   \includegraphics[width=7in]{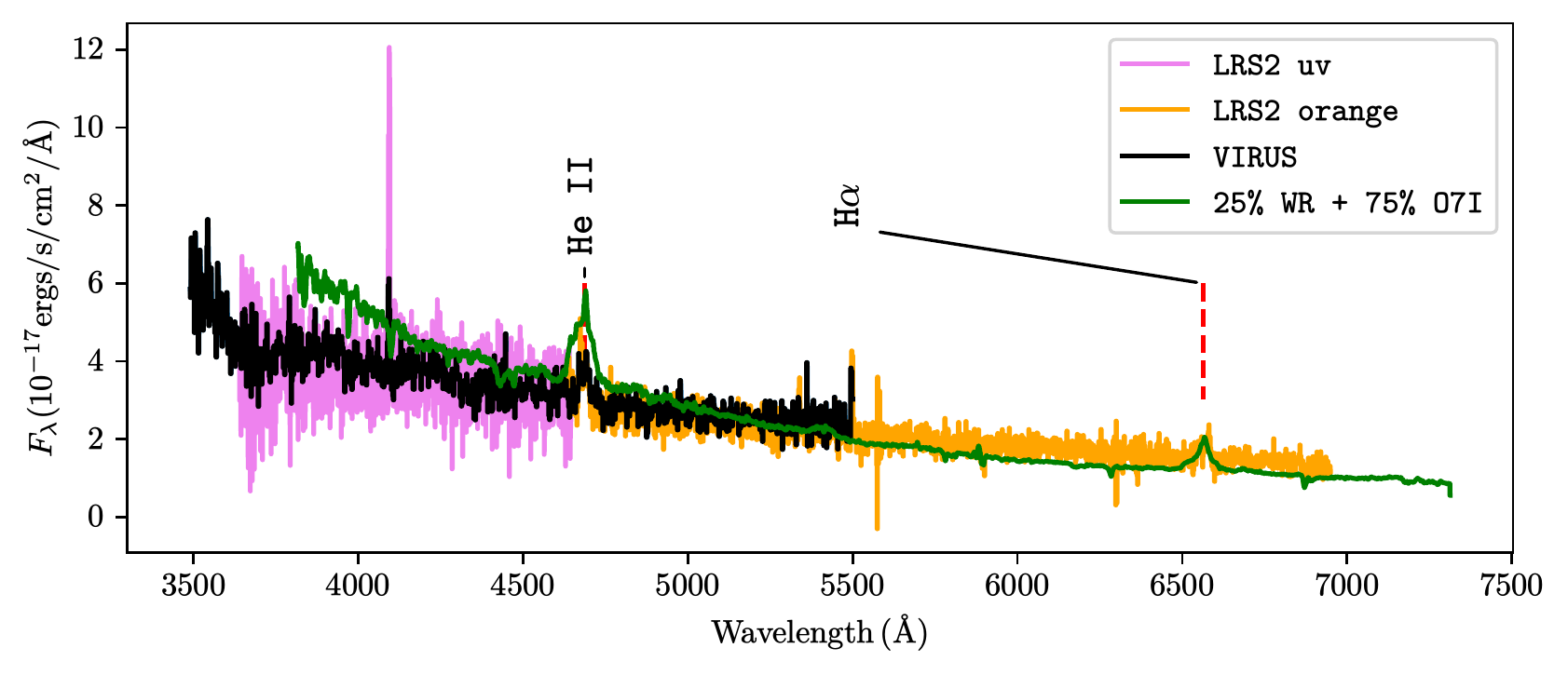}
      \caption{Spectra of ZTF19abvionh.
      The narrow [\OII{}] 
      $\lambda3727$
      is from a galaxy at redshift $\sim$0.1 that is near to ZTF19abvionh on the sky, but unrelated.
      The figure shows all the data from VIRUS
      and LRS2-B overlaid with a spectrum of a WN type Wolf-Rayet star (25\%) plus an O7I star (75 \%) at zero redshift. The WR components corresponding to
      [\HeII{}]
      $\lambda4686$ and H$\alpha$ align with features in the spectra of ZTF19abvionh.}
      \label{fig:ZTF19abvionh}
 \end{center}
\end{figure*}

We elected to observe the fourth ZTF candidate, ZTF19abvionh, with VIRUS on 9/2/19. We detected a nearly featureless continuum with narrow emission lines corresponding to [O II] $\lambda$3727 and $\lambda$5370 and H$\beta$ at a redshift of ~0.1 from a galaxy near to ZTF19abvionh on the sky. That galaxy, GALEXASC J165500.03+140301.3, was about 2.5$\sigma$ more distant than the estimated distance of the merger candidate. 
We submitted a GCN \citep{roselle19e} at 12:05 am and another \citep{roselle19f} at 12:39 am after estimating a black body temperature of about 10,500 K. 

The next night, 9/3/19, we observed ZTF19abvionh with both VIRUS and LRS2-B. These data showed broadened emission features at about 4686\AA\ and 6560\AA\ as shown in Figure~\ref{fig:ZTF19abvionh} that were compatible with He II $\lambda4686$ and H$\alpha$ (or maybe He II $\lambda$6560) at a negligible redshift. This spectrum is compatible with a Wolf-Rayet star of type WN at $\sim 1$ Mpc. Since we saw no host galaxy, the host would have to be a previously unknown dwarf galaxy \citep[e.g.,][]{mcquinn20} accidentally along the line of sight, but unrelated to GALEXASC J165500.03+140301.3. 

We checked an alternative hypothesis that ZTF19abvionh was an Ultra-luminous X-ray Source (ULX) that can have similar spectra, but could be at much greater distances, $\sim 10$ Mpc, by requesting a {\textit{Swift}} observation. 
{\textit{Swift/XRT}} observed the field on 2019-09-20 with an exposure of 3921 s and on 2019-09-23 with an exposure of 435 s. The source was not detected in either observation. We combined the two exposures and used a source region of 45\arcsec\ radius centered on 16:55:00.212, +14:03:04.67. In this region, there are six counts in the 0.2-10 keV band. In a nearby source-free region of 5\arcmin\ radius, there are 251 counts in 0.2-10 keV, for an average of 0.000888 background counts per square arcsecond, or 5.65 in our source region.  Using the Bayesian method of \citet{Kraft91}, we calculated a 90\% confidence upper limit of 5.67 counts, for a countrate of 0.0013 c/s. We assumed an absorbed power-law spectrum with photon index of two and column density of $nH = 4.81 \times 10^{20}$ cm$^{-2}$ \citep{HI4PI}. We used WebPIMMS to calculate a 90\% upper limit of $5.6 \times 10^{-14}$ erg~cm$^{-2}$~s$^{-1}$ in the 0.2-10 keV band.

The lack of detection could be because the source is a WN star at 1 Mpc, but could also be because ULX are variable, and the source had evolved to an X-ray minimum. ZTF19abvionh then underwent solar occlusion. We submitted a third GCN specifically concerning ZTF19abvionh \citep{roselle19g}.

\section{Quantitative Assessment of Null Results}
\label{sec:assess}

While a careful visual inspection of all the data cubes obtained by the LIGHETR program has been performed and no obvious electromagnetic counterpart has been found, the need for a quantitative error on our assessment is still necessary.
For this purpose, we performed two different subsequent sets of analysis on our data, one using the standard continuum grid search and point source extraction routines provided by HETDEX \citep[described in full in][]{gebhardt21} and one using an adaptation of \textsc{scarlet} \citep{scarlet}, a package that performs source separation on multiband images designed for the Legacy Survey of Space and Time (LSST) Science Pipeline.
Our preliminary investigation of both methods highlights the challenges of making quantitative estimates to our sensitivity to OTs. 
We now discuss these analyses and provide a prescription for a path forward. 

\subsection{Point Source Detection and Extraction via HETDEX's Standard Pipeline}
\label{point}

Most of LIGHTER's target fields are centered, by the nature of the search, on extended sources (galaxies), with the occasional centering on an externally detected transient. Although HETDEX's tools are not designed to deal with extended sources and deblending, they do routinely perform point
source detection and extraction. Under optimal conditions, they can detect
point sources up to magnitude 23 in the g-band \citep{gebhardt21}. While the one datapoint at hand, AT2017gfo, did appear as a point source in the outskirts of its host galaxy, uncertainties in the intensity of supernova kicks and the orbital energy dissipated in the explosion debris leave unclear whether kilonovae are to be preferentially found in the outskirts of galaxies. Even if that were to be the case, it is likely that many will be placed within the line of sight of the continuum of the galaxy. With this caveat in mind, we briefy present HETDEX's standard continuum grid search and point source extraction techniques and the results of applying them to our dataset to assess the existence or absence of a transient within our fibers.

For the continuum search, spectra from a VIRUS data cube are flagged as possible continuum sources if they have at least 0.5 ergs s$^{-1}$ cm$^{-2}$ \AA$^{-1}$  either in the blue (from 3700 to 3900 \AA) or in the red (from 5100 to 5300 \AA). Each possible continuum source then gets searched around a 1.5\arcsec\ $\times$ 1.5\arcsec\ grid of 0.1\arcsec\ $\times$ 0.1\arcsec\ resolution, and the spatial location with the maximum flux is selected as the location of the source. Around each location, a point-spread function (PSF) extraction of 3\arcsec\ radius in aperture is subsequently performed. The particular aperture and specifics of the weights that go into the PSF are determined both by typical seeing conditions and known systematics in the design of the instrument (particularly, regarding lack of an atmospheric dispersion corrector and the fiber dithering pattern by which we fill in our collected area). With a set of possible continuum sources in hand for the fields around all of our targets, we compare them with archival sources from the Pan-STARRS catalogue.

\begin{figure*}
    \centering
    \includegraphics[width=\textwidth]{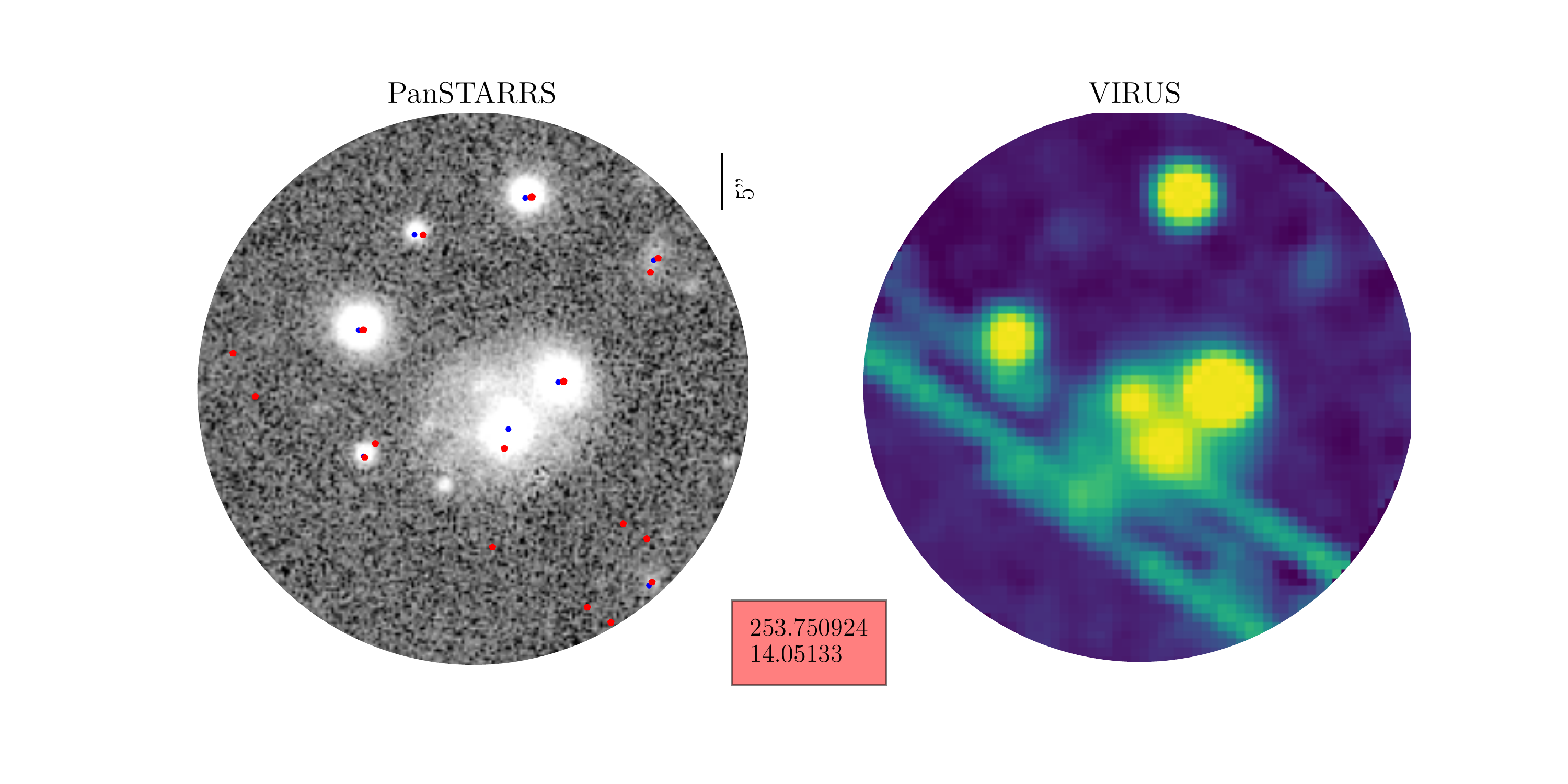}
    \caption{Target galaxy \texttt{GALEXASC J165500.03+140301.3}. On the left is Pan-STARRS imaging of the field overlaid with Pan-STARRS catalogued sources (blue) and VIRUS point sources (red). On the right is the collapsed spectral image of a VIRUS shot of the same field cropped around a 25\arcsec\ radius. The streaks across the IFU field correspond to CCD artifacts at the time of the observations and affected point source identification. This artifact are primarily hot pixels and charge traps that have been subsequently identified and removed from the latest reductions.}
    \label{fig:imagecomp}
\end{figure*}

\begin{figure}
    \centering
    \includegraphics[width=\linewidth]{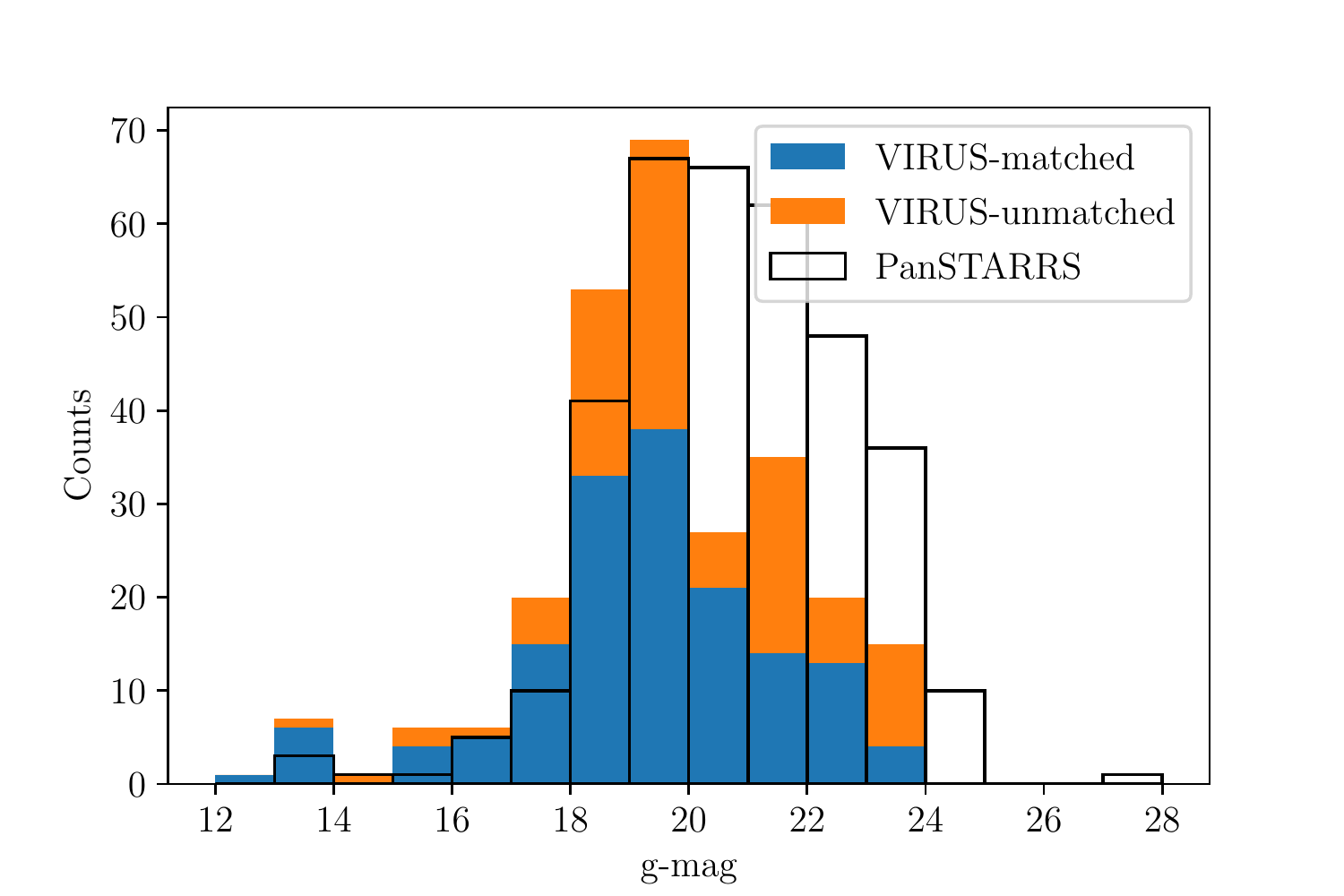}
    \caption{Point source magnitudes for all of our target fields. Transparent bins correspond to Pan-STARRS catalogued sources. Blue and orange are stacked and correspond, respectively, to VIRUS sources matched with Pan-STARRS or only detected by VIRUS. The point source identification algorithm fails with the brightest targets, where saturation leads it to identify more than one source per target. For lower brightness, the mismatch can be explained by spurious continuum point source identification both from Pan-STARRS and VIRUS.}
    \label{fig:maghist}
\end{figure}

Figure~\ref{fig:imagecomp} shows, for reference, one of our target galaxies, \texttt{GALEXASC J165500.03+140301.3}. As can be seen in the image, the CCDs had artifacts across a large fraction of the IFU's footprint. These artifacts made for spurious point source detections and hid some catalogued sources from detection. The problem has since been fixed with improved reductions. Extended sources like galaxies or effectively extended sources like bright stars were also tagged as more than one source, due to oversaturation and to the point source algorithm failing to identify them as single sources. Figure~\ref{fig:maghist} shows a binned histogram of our point source detections for all of our targets. Point sources detected by VIRUS that are less than 2\arcsec\ from a Pan-STARRS source, with a g-band magnitude difference of less than 1 magnitude, are considered coincident. The histogram shows over-identification of bright sources and under-identification of low brightness ones. Visual inspection of our targets shows this is mostly due to saturation and spurious point source identification in extended sources both from Pan-STARRS and VIRUS. While we expected the identification of points sources within the continuum of the galaxy to have significant difficulties, the large number of misidentifications outside of that continuum both for bright and dim sources makes it impossible for us to place limiting magnitudes on point sources for the fields observed. We thus turn to an altogether different detection algorithm for our search.

\subsection{Non-parametric extended source identification with {\sc scarlet}}
\label{scarlet}

Many of the difficulties of our search with VIRUS data cubes and the standard pipeline are due to the intrinsic spatial unevenness of a dithering IFU. The fibers fall in different places of the CCD and have inherent flux variations that can be as large as a factor of $\sim$ 2. While {\sc remedy} does a superb job on the weighting of this variation, software can only do so much and will never be comparable with direct imaging in terms of spatial evenness. The power of spectral data cubes is in that extra spectral dimension and a tool that fully uses this advantage should be employed.

It is in the context of these difficulties that {\sc scarlet}, a software tool
designed for multiband source separation for the LSST Science Pipeline, appears to be a promising alternative \citep{scarlet}. {\sc scarlet} makes use of a constrained matrix factorization in which each source is modeled with a Spectral Energy Distribution (a spectrum) and a non-parametric morphology (the equivalent of
a PSF). The code also allows the imposition of priors on the shape of each
source, allowing further constraints. {\sc scarlet} seems particularly suitable for our problem. The algorithm can make use of the spectral information to identify neighboring fibers contributing to the same flux, so systematics in the PSF extraction could be significantly mitigated. In addition, artifacts could also be removed from the CCD, since again the bleeding into adjacent pixels should show similar spectral features.

\begin{figure}
    \centering
    \includegraphics[width=\linewidth]{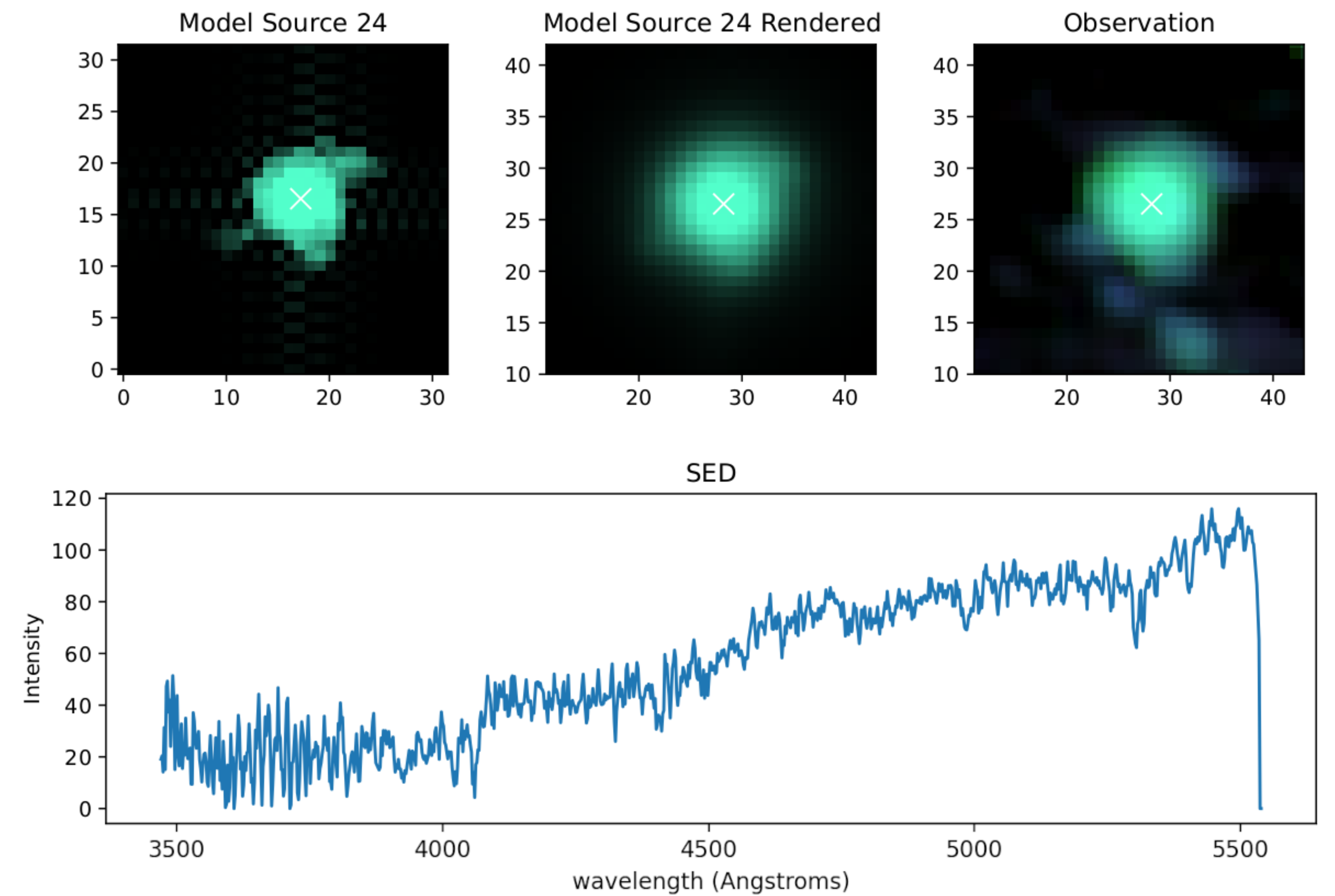}
    \caption{Preliminary results of an adaptation of \textsc{scarlet} to work on VIRUS spectral cubes. Images are on the same scale but with inherent different resolutions and displayed in pixel space. Upper left shows one of the extended sources modeled by \textsc{scarlet}. Upper middle shows the source, rendered to the given PSF. Upper right shows the actual observation compressed along spectra. The colors have been mapped to the corresponding wavelengths of maximum intensity. In the bottom is a picture of the overall collected spectra. Results are promising but need to be combined with a customized continuum grid search routine to be usable.}
    \label{fig:scarlet}
\end{figure}

Figure~\ref{fig:scarlet} shows an individual source from a VIRUS observation extracted by \textsc{scarlet}. 
The package treats point sources, on which a symmetry prior is enforced, or multiple overlapping sources with non-parametric morphologies. 
Constraints on the shape of the spectra can also be enforced. The ability to use analytic priors to fit against, say, a black-body spectrum, were implemented recently in \textsc{scarlet} due to a suggestion from MJRB.
One of the major limitations on the immediate use of \textsc{scarlet} for our purposes is that it requires approximate positions for the sources to be given beforehand.
As shown in Figure~\ref{fig:imagecomp}, in its current state the standard continuum grid search when applied to our data suffers from too high a degree of mismatch to be usable as a source of positions. 

Given these difficulties, our final approach was to simulate a set of point sources with a black body spectrum and place them randomly within a VIRUS data cube to assess under what conditions could they be recovered. We performed a maximum likelihood analysis fit on the data cube, with which we matched a PSF-weighted 3D source (2D for space, 1D for spectra) to all positions in the data cube that hosted the injected sources. We observed that we could only detect sources down to a few magnitudes brighter than the nominal value quoted for HETDEX's pipeline, but we suspect flux calibration issues were at the heart of this difference. Proper knowledge of the PSF weights accessible to HETDEX point source extraction algorithms will be of extreme value for this method to be usable, and further investigations are ongoing. A combination of the continuum grid-search and this method might also be of great value, since the overidentification of continuum sources could be mended if the non-parametric PSF of {\sc scarlet} engulfs them all. 

\section{Models Guiding Observations and Motivation for Spectral Followup}
\label{sec:models}

Observations of GW170817/AT2017gfo closely confirmed expectations of theoretical models. The merging neutron stars create a tidal tail of material ripped from their surfaces and ejected in the orbital plane. This material is expected to be rich in lanthanides, with very large opacity, strong absorption in the blue, and hence characterized by a red continuum \citep{2017Natur.551...80K, Tanaka17}. 
Orthogonal to the orbital plane, ejecta are expected to be expelled at about 0.3c in a jet or cocoon
of somewhat lower opacity material radiating in the blue. AT2017gfo first showed a blue continuum evolving over the course of a few days into a red continuum \citep{2017Sci...358.1574S, 2017ApJ...848L..32M, 2017Natur.551...67P, 2017ApJ...848L..18N, 2018MNRAS.474L..71B}. This was interpreted as first seeing the lower opacity material ejected along the orbital axis and then the high opacity, lanthanide-dominated matter from the orbital plane. Even if other events are very similar, the relative proportions of these components may be different and the orientation will surely be different, so spectral observations promise a rich new haul of insight. A neutron star + black hole (NSBH) merger event will be quantitatively and perhaps even qualitatively different \citep{2019MNRAS.486.5289B, 2019MNRAS.485.4404D, 2021ApJ...917...24Z, 2021ApJ...915...69D}. 

Since we expect any lines to be broadened by the fraction of the speed of light velocities at which the ejecta expand, there could be shallow, broad features in the early spectra as well as fast, transient emission lines. There is a slim chance that one portion of the ejecta, a jet, could irradiate another portion of the ejecta, a tidal stream, and yield a photo-ionized region that would briefy produce emission lines. A fortuitous orientation might mean that broadening is minimized, but even the transverse Doppler shift might be an issue in broadening and obscuring lines. Models of kilonovae show that blue emission is sensitive to the radii of the neutron stars, and that particular aspect angles can show broad yet distinct spectral features as well as a very strong dependance of the flux on angle in the optical band. Early spectral observations in the blue, like the ones LIGHETR is particularly designed to perform, promise strong constraints on the nature of the ejecta.

The LIGHETR program sought to integrate modeling constraints by coordinating with the LANL group simulating merger models, especially their spectra. 
The observational properties of kilonovae are subject to many variables, including the morphology of winds, disks, and jets, the composition on a given line of sight, and, crucially, the aspect angle. To illustrate how early and densely-sampled spectra can serve to constrain these variables, we leverage a large, and continually growing, LANL database of kilonova spectra and light-curve calculations. To demonstrate the importance of the spectra, we present the results from two separate studies: a morphology study~\citep{2021ApJ...910..116K} and a new study designed for this paper varying the composition based on yields from detailed post-merger disk calculations~\citep{2019dMillerpt}. The ultimate goal is to be able to query a data base of models in real time as spectra of candidate OTs are obtained. 

Expected kilonova emission is subject to a broad range of uncertainties, both in the ejecta properties (that depend on aspects of both merger and disk-wind calculations) and in the physics and its numerical implementation.  Examples of these uncertainties span all facets of kilonova emission modeling efforts:  e.g. the amount of matter that is ejected in the dynamical phase of the explosion and distributed in a disk still depends sensitively on the numerics and physics~\citep{2022arXiv220707658H} and on the composition; the ejecta angular and velocity distributions are sensitive to physics modeling such as neutrino transport~\citep{2019dMillerpt}; the emission depends on the nuclear network simulations and the  model for energy deposition~\citep{2021ApJ...918...44B}.  The model database is built on a large suite of models using the  the {\it SuperNu} code~\citep{2013ApJS..209...36W,2016ApJ...827..128V}, adapting initial conditions to study a range of uncertainties in the ejecta composition.  This work includes a suite of models varying the ejecta mass~\citep{2021ApJ...918...10W}, composition~\citep{2018MNRAS.478.3298W,2020ApJ...899...24E}, morphology~\citep{2018MNRAS.478.3298W,2021ApJ...910..116K}, energy sources~\citep{2019ApJ...880...22W}, and atomic~\citep{2020MNRAS.493.4143F} and nuclear~\citep{2018ApJ...863L..23Z} physics\footnote{Many of these models are available at \url{https://ccsweb.lanl.gov/astro/transient/transients_astro.html}}.

\begin{figure}[!ht]
   \begin{center}

\includegraphics[width=\linewidth]{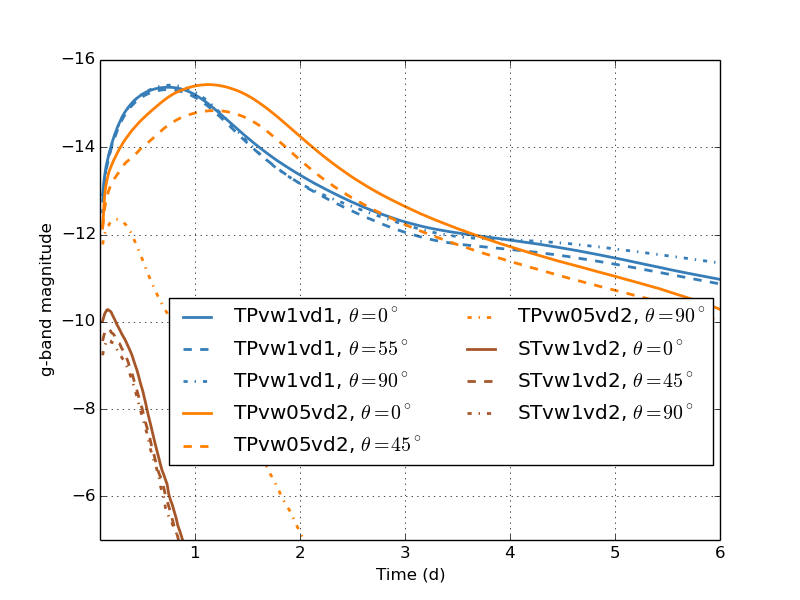}
\includegraphics[width=\linewidth]{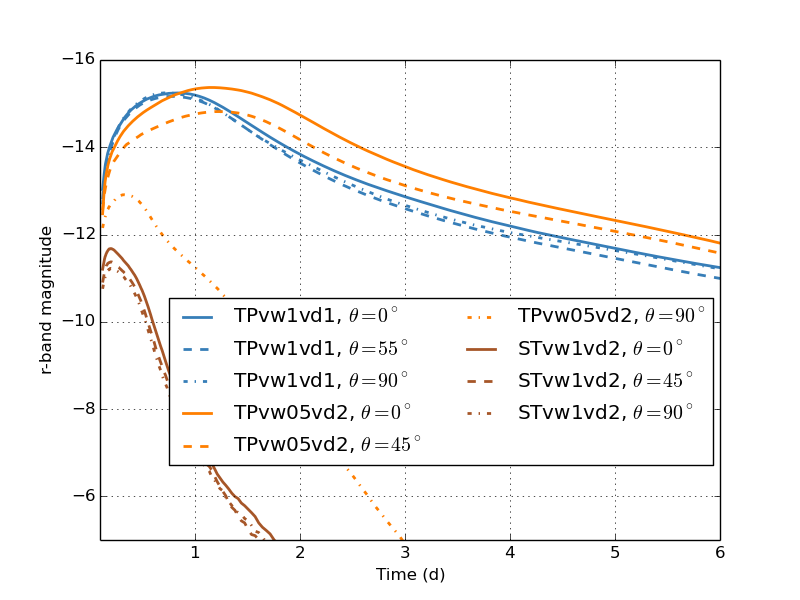}
      \caption{g-band (top) and r-band (bottom) magnitudes of three different two-component models varying morphologies and average ejecta velocities (see ~\cite{2021ApJ...910..116K} for details).  Despite the fact that ejecta masses are the same, the different morphologies and velocities of the components produce different lgiht-curves.  Because of the complex morphologies, the light-curves can also vary with viewing angle.  
}
\label{gband}
 \end{center}
\end{figure}

The significant modeling uncertainties currently make it difficult to quickly and unambiguously identify which transients in the LVC localization maps to follow up with LIGHETR observations.  We hope to use our growing suite of models to help identify the transients to follow-up with LIGHETR and then use LIGHETR observations to further constrain the properties of the kilonova ejecta.  Figure~\ref{gband} shows g- and r-band light curves from models produced by \cite{2021ApJ...910..116K}.  This work varied the morphology, mass, and velocity of a ``wind" (modest electron fraction) and ``dynamical" (neutron rich) component for the ejecta.  The morphologies follow the nomenclature described in the paper including a Toroidal (``T"), Peanut-shaped (``P") and spherical (``S") morphology.  In this figure, we limit ourselves to models with $0.01\,M_\odot$ in each component, but we vary the morphology and average ejecta velocity of each component.  All of the models are characterized by a fast rise (fraction of a day) and decay time (from a fraction of a day to 2 days) for the light-curves.

\begin{figure}[!ht]
   \begin{center}
      \includegraphics[width=\linewidth]{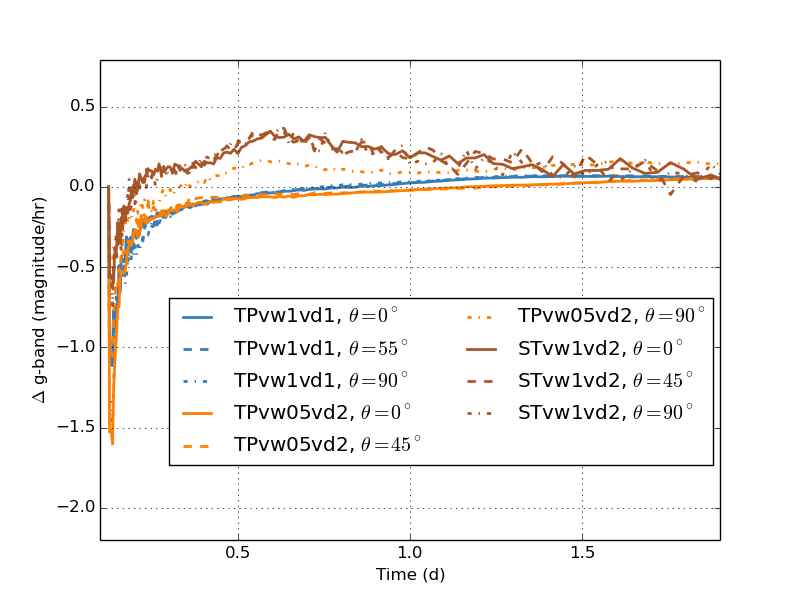}
      \includegraphics[width=\linewidth]{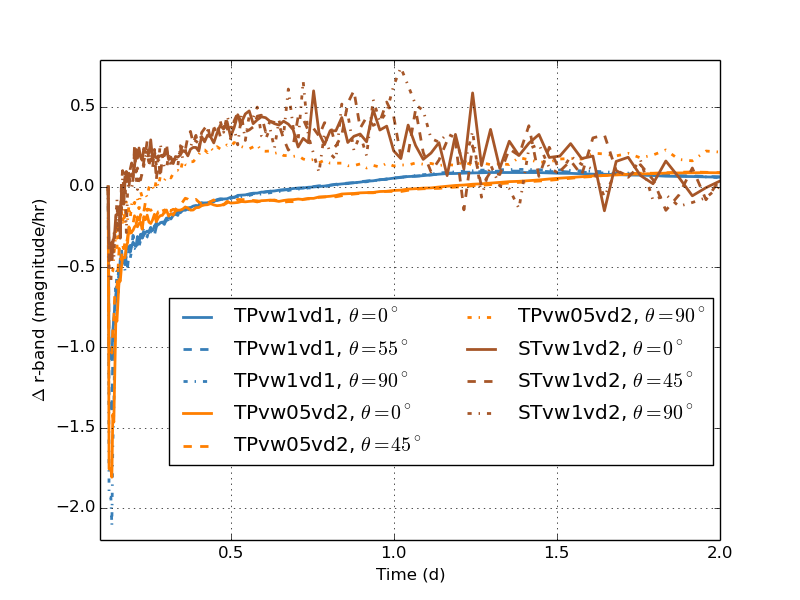}
      \caption{
The rage of change (magnitudes/h) in the g- and r-bands for the models in Figure~\ref{gband}.  All of the models are characterized by a sharp rise in the emission.  But if we miss this initial rise, the evolution can vary dramatically depending on the model.  Within the first 12 hours, some models decline, others continue to rise, but at a slower pace.  All are characterized by rapid variability and, although other transients may exhibit this variability, this rapid variability is a strong indicator of a potential kilonova. 
}

\label{gbandslope}
 \end{center}
\end{figure}

With 2-3 days of observations, the fast evolution in the optical would be a strong indicator of a kilonova observation.  To guide follow-up observations, we need observational indicators already on day 1. In an attempt to quickly differentiate these models, we are investigating a range of discriminating light-curve features. For example, in Figure~\ref{gbandslope} we calculate the change in magnitude with time of the models presented in Figure~\ref{gband}. Most of our light curves are characterized by an initial rapid rise followed by a slower decay.  The rate of this decay depends on the model. In most cases, the rapid evolution all occurs within the first day. After the first 6 hours, some models are already decaying.  Others continue to rise for 12-24 hours.  In all cases, the variability is high in kilonova models and variation of a few tenths of a magnitude in a few hours is an indication of a potential kilonova.

Once we identify potential kilonovae and obtain observations, we plan to use these same simulations to help interpret the observations.  A variety of ejecta yields were predicted from different analyses of AT2017gfo~\citep{2018ApJ...855...99C}.  The results of studies of potential kilonovae associated with gamma-ray bursts have also shown a variety of interpretations for a single data set~\citep{2021ApJ...906..127F,2021MNRAS.502.1279O}.  Constraining this range of interpretations of a single data set requires a broad set of observations and the HET can play a critical role in differentiating model interpretations.  Differentiating the models can be done with light-curve bands, but spectral observations also provide insight into the properties of the ejecta.  For example, although the fast ejecta velocities and dense forest of lines produce continuum-like spectra, both observations of AT2017gfo and a series of models suggest that line features might be observed across a broad spectral range~\citep{watson,2021ApJ...910..116K,domoto22}.

\begin{figure*}[!ht]
   \begin{center}
\includegraphics[width=3.5in]{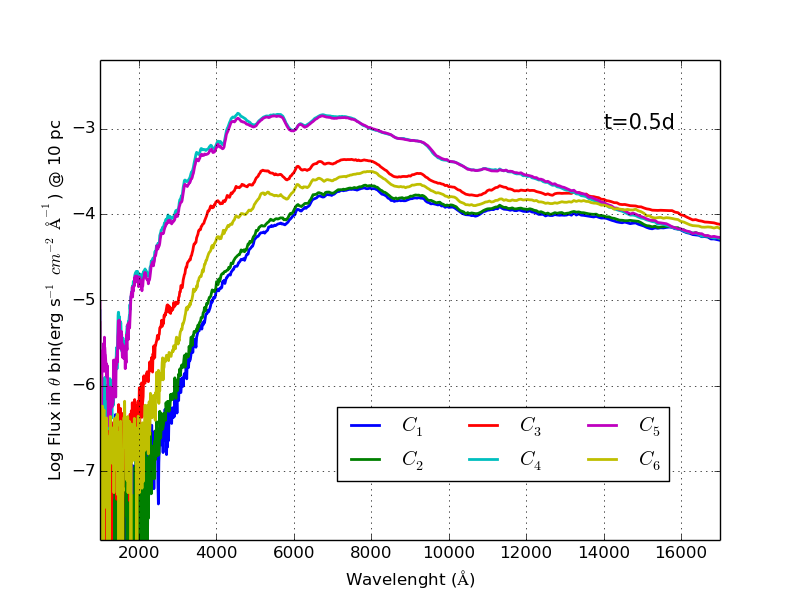}
\includegraphics[width=3.5in]{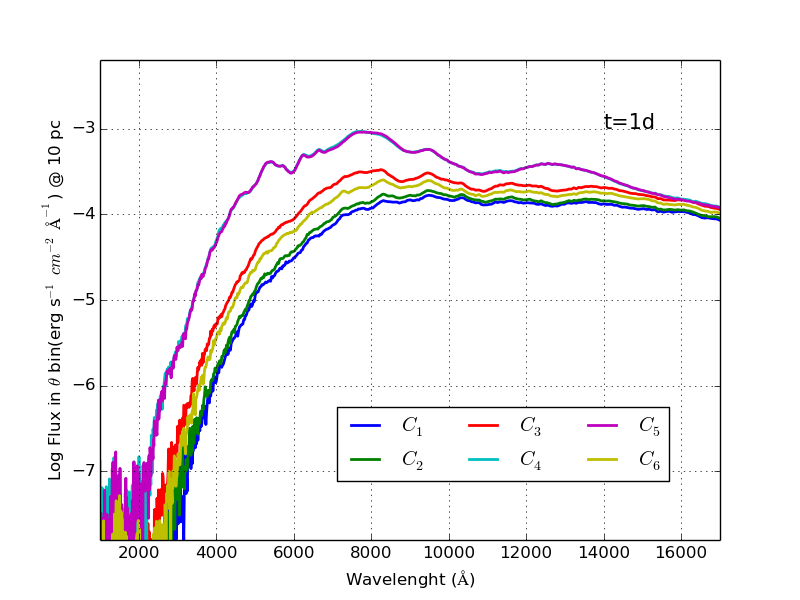}
\includegraphics[width=3.5in]{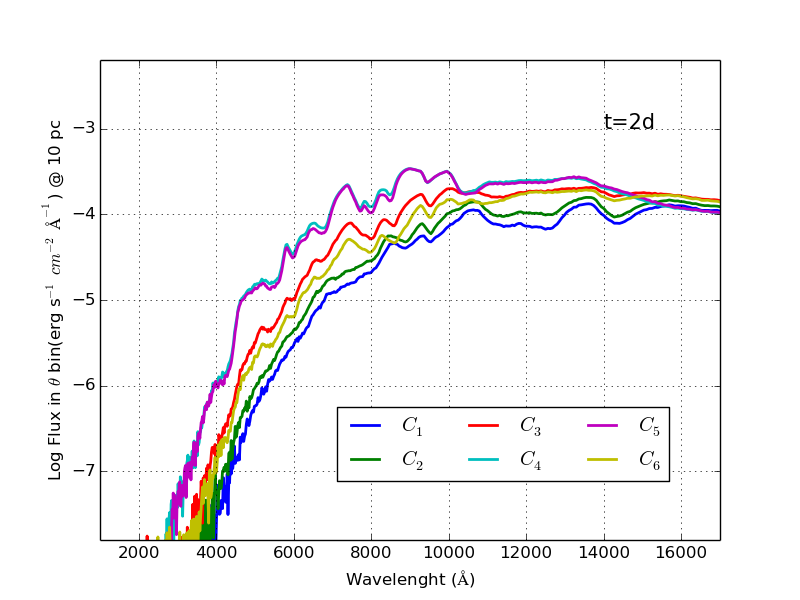}
\includegraphics[width=3.5in]{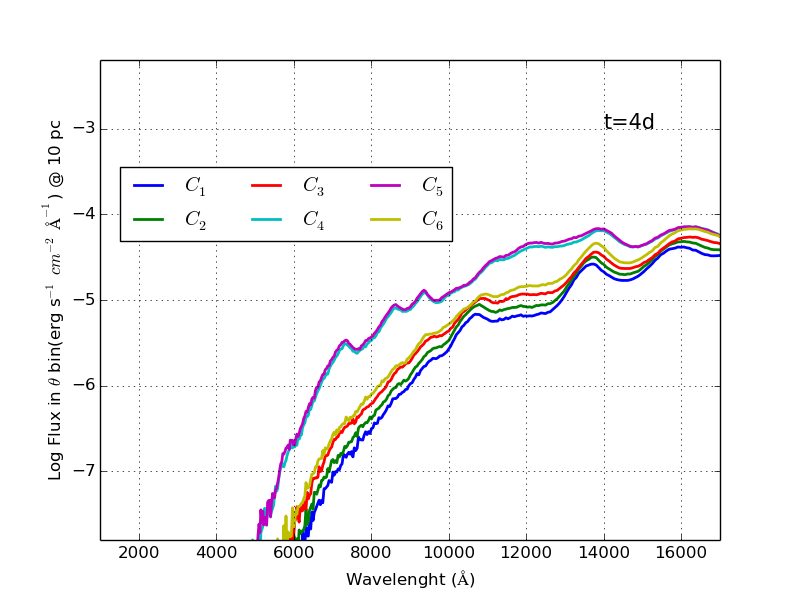}
      \caption{Spectra from a set of compositions produced by a high-fidelity spherical wind-ejecta model at a variety of times for a range of angles using the wind disk model from~\cite{2019dMillerpt} at 0.05, 1, 2, and 4 days.  Even though these results are focused on a single wind profile, varying the abundances produce a range of spectral fluxes, especially in the optical bands that vary by over an order of magnitude.  The compositions studied in this plot are listed in Table~\ref{tab:composition}. 
      }
      \label{specex}
 \end{center}
\end{figure*}

Many of these spectral features are difficult to detect, requiring high signal-to-noise spectra, and often-times these features are line blends from multiple elements. Despite these difficulties, broad line features and spectral slopes can help us constrain the ejecta composition. Broad spectral features can also probe the morphology and viewing angle of the merger.  Figure~\ref{specex} shows the spectra at four epochs in time in a series of as-yet unpublished spherically-symmetric wind models using ejecta compositions spanning a range of distributions.  The compositions in Figure~\ref{specex} are listed in Table~\ref{tab:composition}.  Compositions $C_1$ and $C_2$ have high mass fractions of heavier elements and display relatively few features beyond some broad features above $8000$~\AA\ at two days.  As we decrease the fraction of these heavy elements ($C_4$ and $C_5$ have the lowest mass fractions), a number of spectral features from elements up to the first r-process peak are visible.  The wide variation in this spherically-symmetric wind model, particularly in the optical bands, demonstrates just how sensitive the HET spectra could be to the ejecta characteristics. We can also use these broad spectral features to help us confirm an electromagnetic counterpart, allowing us to notify the community of a true kilonova detection.

\begin{figure}[!ht]
   \begin{center}
\includegraphics[width=\linewidth]{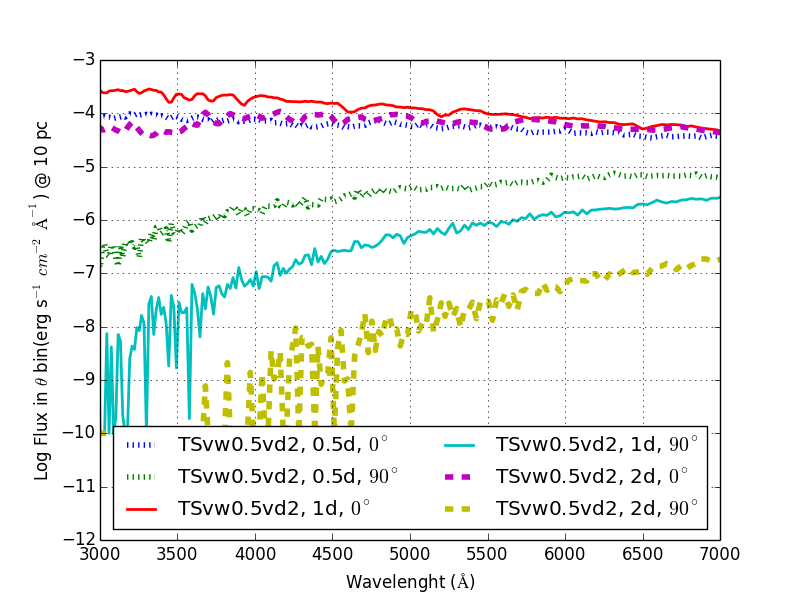}
      \caption{
Model spectra at 0.5, 1.0, and 2.0 days for a model TSvw0.5vd2 from~\cite{2021ApJ...910..116K} along 2 different viewing angles (along the angular momentum axis and perpendicular to this axis).  }
\label{spectime}
 \end{center}
\end{figure}

Composition is just one of the properties of model kilonovae that we have studied.  Figure~\ref{spectime} shows the spectra in the 3000--7000~\AA\ wavelength range focusing on one of the models from the morphology study~\citep{2021ApJ...910..116K} with a spherical wind of steady velocity 0.5 c and a toroidal disk component with a velocity of 0.2~c (TSvw0.5vd2), comparing spectra at different times and different viewing angles using the standard ``wind 1" composition from~\cite{2021ApJ...910..116K}.  The variation with respect to viewing angle for this particular morphology is dramatic. Without constraining the viewing angle, we have yet to identify distinct model spectral features in the early blue kilonova spectra.  We do expect any line features to be broad and, as the LANL team improves its opacities, we will continue to look for specific spectral features.  These features are essential to help us distinguish between the morphology, composition, and viewing angle effects.  Even so, extensive observations are required to successfully disentangle these effects, and HET spectra can be a critical part of this effort.

Our model database continues to grow as we incorporate new physics and initial conditions into our models and import the simulations of other groups into our studies.  The kilonova community is conducting comparison projects to better constrain issues in physics and numerics. These studies also will ultimately allow us to do better uncertainty quantification as well.

\section{Future Observations}
\label{sec:future}

The LVC third observing run brought two new factors into play. The greater sensitivity increased the search volume thus increasing the likelihood of detecting a BNS event. This also meant, however, that the average merger was more distant and the OT more difficult to detect. It is also true that the factors influencing the intrinsic luminosity and color of a BNS kilonova are many, especially including the aspect angle, so it may not be reliable to scale expectations with the observations of AT2017gfo. The bottom line is that no group anywhere detected an OT in O3, despite massive deployment of resources. That said, we are confident that our original analysis of the capabilities of the HET stands, and that the program merits continuing with the same commitment of HET resources.

For the O4 observing run, {\sc diagnosis} will listen to 
alerts sent by LVK on Kafka Notices via SCiMMA\footnote{https://rtd.igwn.org/projects/userguide/en/stable/quickstart.html}, which is the method by which LVK alerts will be sent in O4. In addition, galaxies within the region observable by HET will be weighted more accurately by mass using the updated GLADE catalog.

LIGHETR collected a considerable amount of ancillary data during O3. We will make these data available upon request. 
We will make our O3 pointings available to the Treasure Map\footnote{http://treasuremap.space/} record of O3 data and intend to implement deposition of our real time pointing record in O4.

The O3 science run drove home the difficulty in identifying the kilonova resulting from a compact object merger among the extensive lists compiled by optical astronomers of contemporary transients that fell within the gravitational wave localization errors. To help us focus on the most-likely candidates to follow-up with our telescope time, we plan to leverage a large, and continually growing, database of model kilonova spectra and light-curve calculations from our LANL collaborators. These simulations can be used both to guide our observations by helping us determine which transients to follow-up and to better interpret our results once observations are made.

More work remains to be done to strengthen the connection between models and the observational data. Kilonovae templates make use of a broad suite of models from the LANL simulation effort, with physics uncertainty studies covering opacity, composition, power source and morphology. Template matching to these models could improve our sensitivity to low brightness kilonovae and also help in the effort of distinguishing kilonovae from other transients in real time, which could significantly reduce the amount of telescope time the community devotes to LVC alerts. When and if a source were to be detected by LIGHETR, we could also wield the templates to make inferences on the kilonova properties and thus provide the community with a more detailed picture of physics behind the transient.

The pipelines we are developing for LIGHETR are designed to incorporate the growing data base of LANL models. We have recently incorporated a neural network-based classifier that scans each pixel in an IFU data cube and can identify kilonova candidates by comparison to theoretical models.  This classifier was trained on the LANL suite of models and will be updated with any new theoretical developments. During an active search, the software can spectroscopically identify a kilonova candidate and then rapidly compare to the LANL model grid to constrain model parameters such as the critical aspect angle. We will thus have a substantial theoretical and simulation effort to complement our observational program.

The challenges faced by both the standard HETDEX pipeline and {\sc scarlet} in extracting candidate OTs from the VIRUS data illustrate the value of repeated observations of our target galaxies. With repeated observations we could perform data cube differentiation and any transients, be they systematic or real, would be identified and characterised. This would also help assessing the performance of our methods, since we could compare our deblended extraction
of some of the transients we followed-up with the differentiated data. 

While it is clear much needs to be done to obtain a quantitative figure of merit for non-detections, we should also highlight the great success of the LIGHETR program. The LIGHETR Collaboration achieved the production of an effective alert system for possible kilonovae transients, {\sc diagnosis}, customized to the specifics of our instrument. The creation of an extremely quick and powerful reduction pipeline, {\sc remedy}, has also been successfully achieved for the purposes of this endeavour. These tools allow members of the collaboration to have, within minutes of a gravitational wave alert, access to spectral data cubes of galaxies that have a finite probability of hosting the transient. During O3, these data cubes were obtained, reduced, and immediately inspected for possible continuum sources but no evident kilonovae transients were found.

The data and tools collected during this campaign are of great scientific value regardless of the identification of a transient. 
Kinematic maps for several galaxies have been collected and already {\sc remedy} has become a
standard reduction tool for the HETDEX consortium. Some of the transients we followed up motivated new observing proposals. The Wolf-Rayet star we suspect to be hosted by a low-brightness dwarf galaxy remains a puzzle we hope to solve. 

For O4, the VIRUS array is fully implemented with 78 VIRUS units with a total of 34,944 fibers spanning an area 21\arcmin\ on a side (Figure \ref{fig:hetdex}). 
Previous VIRUS operations have shown that typical set up time is about 6 minutes. An exposure of 9 minutes will get down to about 19th magnitude. Although in this project the average time for set up was only 2.3 minutes, we still plan on a single exposure on a target galaxy to require a total time of 15 minutes. There will be a great deal of competition from wide-field robotic photometric facilities to discover the OT, but if we find the OT first, we will automatically acquire the first spectrum. Even if we do not find the OT first, by concentrating on a high confidence region of the candidate skymap HET will be roughly positioned quickly to get early spectra. If we see an early blue component, as widely expected, that will already be interesting. If we do not see an early blue component, that will also be interesting. In any case, the blue response of the HET VIRUS array is well designed to get this early data in the blue.

\newpage

\section*{Acknowledgments}
\label{sec:ackn}

We thank Gary Hill for discussions of the response functions of VIRUS and LRS-2. 
We thank Chad Hanna and David Radice for discussions in the initial stages of forming the LIGHETR program.

BPT and JCW are supported in part by NSF grant 1813825, by a DOE grant to the Wooten Center for Astrophysical Plasma Properties (WCAPP; PI Don Winget), and by grant G09-20065C from the Chandra Observatory.
KG acknowledges support for this work from NSF-2008793.
AZ was supported by NSF Grant Numbers PHY-1912578 and PHY-2207594 during this work.
RM acknowledges the intellectual support of The Weinberg Institute, The Center for Gravitational Physics, and the Department of Physics at The University of Texas at Austin.
DP is supported in part by the National Aeronautics and Space Administration through Chandra Award Numbers GO0-11007A and GO GO9-20065A issued by the Chandra X-ray Center, which is operated by the Smithsonian Astrophysical Observatory for and on behalf of the National Aeronautics Space Administration under contract NAS8-03060.
JV is supported by the project “Transient Astrophysical Objects” GINOP 2.3.2-15-2016-00033 of the National Research, Development and Innovation Office (NKFIH), Hungary, funded by the European Union.
Time domain research by the University of Arizona team and D.J.S.\ is supported by NSF grants AST-1821987, 1813466, 1908972, \& 2108032, and by the Heising-Simons Foundation under grant \#2020-1864.
The work by CLF, OK, and RW was supported by the US Department of Energy through the Los Alamos National Laboratory. Los Alamos National Laboratory is operated by Triad National Security, LLC, for the National Nuclear Security Administration of U.S.\ Department of Energy (Contract No.\ 89233218CNA000001). 

The University of Texas at Austin and McDonald Observatory sit on indigenous land. The Tonkawa lived in central Texas and the Comanche and Apache moved through this area. The Davis Mountains that host McDonald Observatory were originally husbanded by Lipan Apache, Warm Springs Apache, Mescalero Apache, Comanche and various tribes of the Jumanos. We acknowledge and pay our respects to all the Indigenous Peoples and communities who are or have been a part of these lands and territories in Texas. 

This research was made possible by the open release of gravitational wave candidates detected by the Advanced LIGO and Virgo observatories.
LIGO Laboratory and Advanced LIGO are funded by the United States National Science Foundation (NSF) as well as the Science and Technology Facilities Council (STFC) of the United Kingdom, the Max-Planck-Society (MPS), and the State of Niedersachsen/Germany for support of the construction of Advanced LIGO and construction and operation of the GEO600 detector. Additional support for Advanced LIGO was provided by the Australian Research Council. Virgo is funded, through the European Gravitational Observatory (EGO), by the French Centre National de Recherche Scientifique (CNRS), the Italian Istituto Nazionale di Fisica Nucleare (INFN) and the Dutch Nikhef, with contributions by institutions from Belgium, Germany, Greece, Hungary, Ireland, Japan, Monaco, Poland, Portugal, Spain. The construction and operation of KAGRA are funded by Ministry of Education, Culture, Sports, Science and Technology (MEXT), and Japan Society for the Promotion of Science (JSPS), National Research Foundation (NRF) and Ministry of Science and ICT (MSIT) in Korea, Academia Sinica (AS) and the Ministry of Science and Technology (MoST) in Taiwan.
 
 \facilities{
This study employs observations obtained with the Hobby-Eberly Telescope, which is a joint project of the University of Texas at Austin, the Pennsylvania State University, Ludwig-Maximilians-Universit{\"a}t M{\"u}nchen, and Georg-August-Universit{\"a}t G{\"o}ttingen.
The HET is named in honor of its principal benefactors, William P. Hobby and Robert E. Eberly. The Low Resolution Spectrograph 2 (LRS2) was developed and funded by the University of Texas at Austin McDonald Observatory and Department of Astronomy and by Pennsylvania State University. We thank the Leibniz-Institut f{\"u}r Astrophysik Potsdam (AIP) and the Institut f{\"u}r Astrophysik G{\"o}ttingen (IAG) for their contributions to the construction of the integral field units. This study also utilized X-ray data from the {\it Neil Gehrels Swift Observatory}, and \chandra.}

\software{
astropy \citep{2013A&A...558A..33A,2018AJ....156..123A}, emcee \citep{2013PASP..125..306F}, numpy \citep{harris2020array}, scipy \citep{2020SciPy-NMeth}, matplotlib \citep{Hunter:2007}, pandas \citep{reback2020pandas,mckinney-proc-scipy-2010}.
\swift\ data were reduced with {\tt XRTDAS} (v0.13.5), {\tt CALDB} (v20190910), {\tt XRTPIPELINE} and {\tt XSELECT}. \chandra\ data were processed with {\tt SPECEXTRACT}.
}


\bibliographystyle{mnras}

\bibliography{lighetr}

\newpage

\appendix
\counterwithin{figure}{section}
\counterwithin{table}{section}

\section{Model Lightcurve Compositions}
\label{appsec:comp}

Table \ref{tab:composition} gives the compositions employed in 
Figure~\ref{specex}.

\begin{table*}
\begin{center}
\begin{tabular}{l|cccccc}
\hline\hline              
Z & $C_1$ & $C_2$ & $C_3$ & $C_4$ & $C_5$ & $C_6$ \\
\hline
24 &  0.0         & 7.49370e-04 & 3.84341e-03 & 9.12667e-03 & 1.08047e-02 & 1.50548e-04 \\
26 &  1.09998e-04 & 2.94116e-03 & 1.60909e-02 & 2.93123e-02 & 4.22312e-02 & 5.53921e-03 \\
34 &  1.20896e-01 & 1.99367e-01 & 2.86043e-01 & 2.92805e-01 & 2.35532e-01 & 1.85846e-01 \\
35 &  3.69738e-02 & 5.20294e-02 & 1.11589e-01 & 2.27298e-01 & 3.27268e-01 & 6.73422e-02 \\
40 &  1.09691e-02 & 2.26226e-02 & 6.74217e-02 & 1.32521e-01 & 1.65283e-01 & 1.69854e-01 \\
46 &  1.47586e-01 & 1.39364e-01 & 1.05772e-01 & 9.16517e-02 & 7.82567e-02 & 1.68748e-01 \\
52 &  6.04172e-01 & 5.22899e-01 & 3.86228e-01 & 2.14772e-01 & 1.38436e-01 & 3.76756e-01 \\
57 &  1.96323e-03 & 1.39742e-03 & 7.12728e-04 & 1.26050e-04 & 1.12861e-04 & 9.23346e-04 \\
58 &  2.95243e-03 & 2.23794e-03 & 1.14957e-03 & 1.81696e-04 & 1.91118e-04 & 1.54047e-03 \\
59 &  7.70276e-04 & 5.68369e-04 & 3.06602e-04 & 1.19901e-04 & 0.0         & 3.27700e-04 \\
60 &  1.92511e-03 & 1.32451e-03 & 5.91964e-04 & 1.19743e-04 & 1.26560e-04 & 1.09535e-03 \\
62 &  2.38889e-03 & 1.58250e-03 & 7.01276e-04 & 1.19743e-04 & 1.15225e-04 & 9.87945e-04 \\
63 &  5.97838e-04 & 3.86548e-04 & 1.78528e-04 & 0.0         & 0.0         & 2.69349e-04 \\
64 &  5.82899e-03 & 3.87802e-03 & 1.72436e-03 & 1.91844e-04 & 1.31481e-04 & 1.05073e-03 \\
65 &  1.54890e-03 & 1.02126e-03 & 4.51920e-04 & 0.0         & 0.0         & 3.63695e-04 \\
66 &  1.31050e-02 & 8.49339e-03 & 3.89174e-03 & 4.32027e-04 & 2.87312e-04 & 2.10961e-03 \\
67 &  2.90705e-03 & 1.87583e-03 & 8.73047e-04 & 0.0         & 0.0         & 4.56880e-04 \\
68 &  9.80968e-03 & 6.47393e-03 & 2.92234e-03 & 3.28006e-04 & 2.32171e-04 & 1.71006e-03 \\
69 &  1.44669e-03 & 9.58414e-04 & 4.32732e-04 & 0.0         & 0.0         & 2.95809e-04 \\
70 &  1.84161e-02 & 1.24477e-02 & 5.52962e-03 & 6.75113e-04 & 5.96579e-04 & 4.89435e-03 \\
92 &  1.56322e-02 & 1.73818e-02 & 3.54577e-03 & 3.39432e-04 & 3.94817e-04 & 9.73867e-03 \\
\hline\hline
\end{tabular}
\caption{Composition of the models presented in Figure~\ref{specex}.}
\label{tab:composition}
\end{center}
\end{table*}

\newpage

\section{LIGHETR Alert Components}
\label{appsec:alert}

\begin{figure*}[!ht]
   \begin{center}

\includegraphics[width=\linewidth]{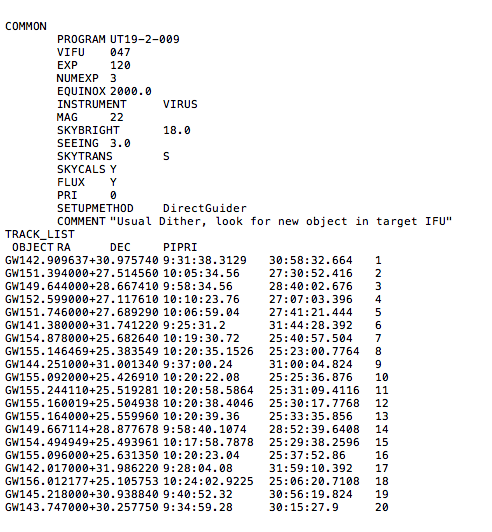}
      \caption{{\scriptsize
Phase II TSL generated by {\sc diagnosis} giving the prioritized list of target galaxies that will be employed to trigger the search for the OT with the VIRUS IFU array. The galaxies in this list are also presented in Figure~\ref{priority}
         }}\label{list}
         
          \end{center}
\end{figure*}

\begin{figure*}[!ht]
   \begin{center}

\includegraphics[width=\linewidth]{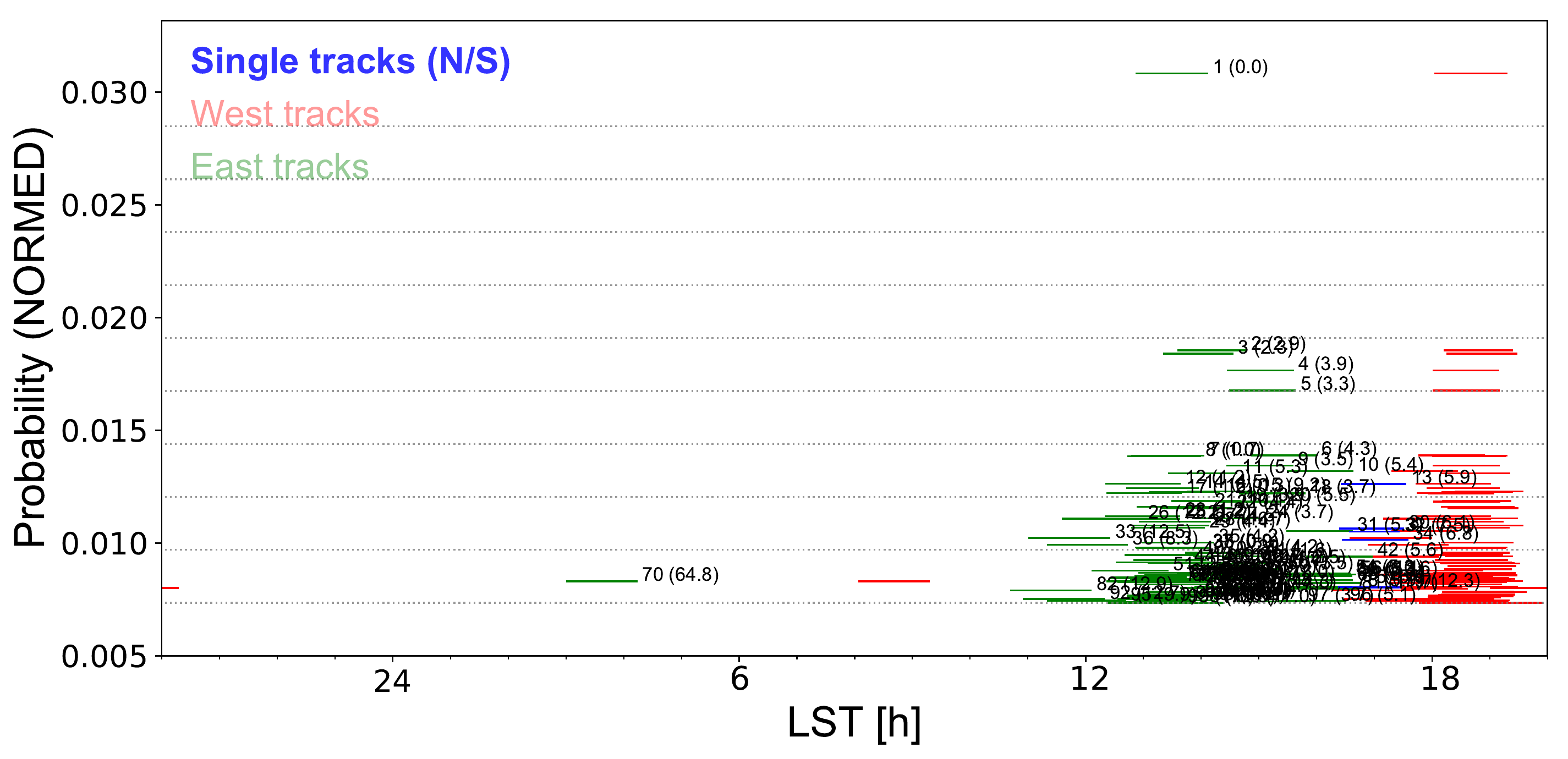}
      \caption{{\scriptsize
Sample output from combining localization information for LIGO event S190425z as illustrated in Figure~\ref{ligohetmap} with the GLADE galaxy catalog to produce a list of 20 galaxies prioritized by the likelihood of being the host of a LIGO alert burst of gravitational waves. The vertical scale gives the log of the probability, and the horizontal 
scale gives the position in hour of right ascension. 
Blue (single bands between the East and West extremes) shows galaxies available on a single track, red (right bands) and green (left bands) show the time when a given galaxy can be observed on the West track and the East track, respectively.
         }}\label{priority}

 \end{center}
\end{figure*}

Here we give some illustrations of how the internal LIGHETR alert system works. Figure~\ref{alert} gives an example of an LVC alert as processed by {\sc diagnosis} for local redistribution. Figure~\ref{list} gives the Target Submission List (TSL) subsequently sent to the Resident Astronomers at the HET. Figure~\ref{priority} presents a graphical representation of the galaxy target priority list used in the original search for an OT.

\begin{figure*}[!ht]
   \begin{center}

\includegraphics[width=\linewidth]{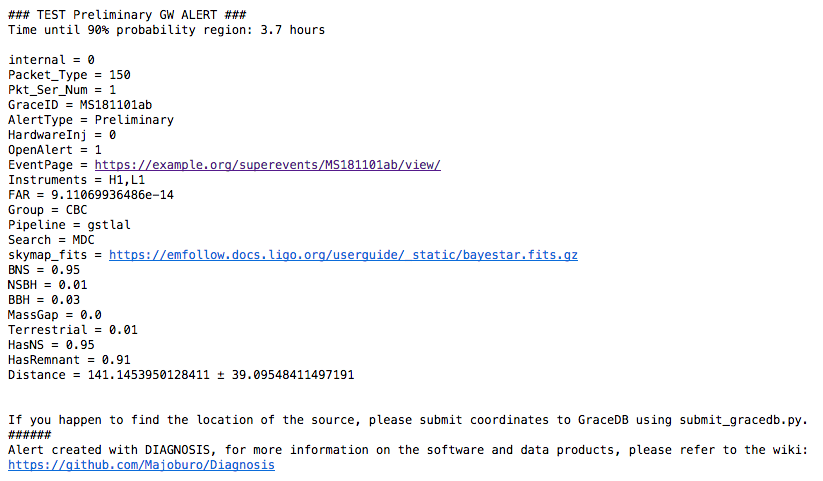}
      \caption{{\scriptsize
Trial LIGO alert as processed by {\sc diagnosis} for local distribution.
         }}\label{alert}
         
          \end{center}
\end{figure*}

\newpage

\newpage

\section{Timetable for Alert, Search, Dense Spectral Sampling, and Analysis}
\label{appsec:time}

{\bf VIRUS} \\
00:00 + 01:00 LIGO alert, precursor or normal, sent to team, RAs \\
01:00 + 00:30 automatically run {\sc diagnosis} tool in response to the alert  \\
01:30 + 00:30  {\sc diagnosis} generates VIRUS Phase II TSL with priority list of galaxy targets \\ 
02:00 + 10:00 scientists in consultation with RAs make decision on whether to trigger telescope \\
12:00 + 02:00 RAs ingest galaxy Phase II into HET queue and decide when to implement, depending on current observation \\ 
14:00 + 01:00 select first galaxy target in HET queue \\
15:00 + 05:00 slew telescope and setup \\
20:00 + 06:00 start first 6 minute dither \\
26:00 + 01:00 run {\sc remedy} code to produce collapsed image and reduced spectrum. Start second dither \\
27:00 + 05:00 inspect {\sc remedy} output. If OT, send alert, submit LRS2 Phase II  \\
32:00 + 06:00 2nd dither out, proceed as above with {\sc remedy}, start 3rd dither \\
38:00 + 05:00 3rd dither out, either move to next target or switch to LRS2 if OT found \\
43:00 -- proceed as above;  start next VIRUS exposure or first LRS2 exposure if OT found \\ 
{\bf Total 180 minutes} \\

{\bf LRS2} \\
\\
First Night \\
00:00 + 05:00 halt VIRUS observations, submit LRS2 Phase II \\
05:00 + 08:00 setup on LRS2-B \\
13:00 + 20:00 first LRS2-B exposure \\
33:00 + 02:00 set up on LRS2-R \\
35:00 + 20:00 first LRS2-R exposure. Reduce and examine first LRS2-B spectrum \\
55:00 + 02:00 setup again on LRS2-B \\
57:00 + 20:00 second LRS2-B exposure. Reduce and examine first LRS2-R spectrum \\
77:00 + 02:00 setup again on LRS2-R \\
79:00 -- continue until target no longer visible for the HET \\
{\bf Total 300 minutes} \\
\\
Second Night \\
00:00 + 08:00 setup on LRS2-B \\
08:00 + 20:00 first LRS2-B exposure \\
28:00 + 02:00 set up on LRS2-R \\
30:00 + 20:00 first LRS2-R exposure. Reduce and examine first LRS2-B spectrum \\
50:00 + 02:00 setup again on LRS2-B \\
52:00 + 20:00 second LRS2-B exposure. Reduce and examine first LRS2-R spectrum \\
72:00 + 02:00 setup again on LRS2-R \\
74:00 + 20:00 second LRS2-R exposure. Reduce and examine second LRS2-B spectrum \\
94:00 + 02:00 setup again on LRS2-B \\
96:00 + 20:00 third LRS2-B exposure. Reduce and examine second LRS2-R spectrum \\
116:00 + 02:00 setup again on LRS2-R \\
118:00 + 20:00 third LRS2-R exposure. Reduce and examine third LRS2-B spectrum \\
138:00 +12:00 Reduce and examine third LRS2-R spectrum \\
{\bf Total 150 minutes} \\
\\
Third, Fourth, Fifth Nights \\
00:00 + 08:00 setup on LRS2-B \\
08:00 + 30:00 first LRS2-B exposure \\
38:00 + 02:00 set up on LRS2-R \\
40:00 + 30:00 first LRS2-R exposure. Reduce and examine first LRS2-B spectrum \\
70:00 + 02:00 setup again on LRS2-B \\
72:00 + 30:00 second LRS2-B exposure. Reduce and examine first LRS2-R spectrum \\
102:00 + 02:00 setup again on LRS2-R \\
104:00 + 30:00 second LRS2-R exposure. Reduce and examine second LRS2-B spectrum \\
134:00 + 06:00 Reduce and examine second LRS2-R spectrum \\
{\bf Total 140 minutes each night} \\
\\
Day 1 -- begin analysis of data, comparison to models \\
Day 5 -- finish analysis of data, comparison to models \\
Day 7 -- submit paper \\

\end{document}